\documentclass[twocolumn]{aastex631}

\usepackage{amsmath} %
\usepackage{bm}

\begin{document}

\title{Multi-wavelength constraints on the local black hole occupation fraction}

\correspondingauthor{Colin J. Burke}
\email{colin.j.burke@yale.edu}

\author[0000-0001-9947-6911]{Colin J. Burke}
\affiliation{Department of Astronomy, Yale University, 266 Whitney Avenue, New Haven, CT 06511, USA}

\author[0000-0002-5554-8896]{Priyamvada Natarajan}
\affiliation{Department of Astronomy, Yale University, 266 Whitney Avenue, New Haven, CT 06511, USA}
\affiliation{Department of Physics, Yale University, New Haven, CT 06520, USA}
\affiliation{Black Hole Initiative, Harvard University, 20 Garden Street, Cambridge, MA 02138, USA}

\author[0000-0003-4703-7276]{Vivienne F. Baldassare}
\affiliation{Washington State University, Department of Physics and Astronomy, WA 99164, USA}

\author[0000-0002-7007-9725]{Marla Geha}
\affiliation{Department of Astronomy, Yale University, 266 Whitney Avenue, New Haven, CT 06511, USA}



\begin{abstract}
The fraction of dwarf galaxies hosting central, intermediate-mass black holes (IMBHs) at low redshifts is an important observational probe of black hole seeding at high redshift. Detections of nuclear accretion signatures in dwarf galaxies provides strong evidence for the presence of these IMBHs. We develop a Bayesian model to infer the black hole occupation fraction assuming a broken power law Eddington ratio distribution function. Our approach accounts for non-detections, incompleteness, and contamination from star-forming-related emission. We apply this model to galaxies with X-ray data from the Chandra Source Catalog at distances $<50$ Mpc, radio data from the VLA Sky Survey at $< 50$ Mpc, and optical variability data from the Palomar Transient Factory at $z<0.055$. We find a black hole occupation fraction of at least $90$ percent at stellar masses of $M_{\star}=10^8~M_{\odot}$ and at least $39$ percent at $M_{\star} = 10^7~M_{\odot}$ (95\% confidence intervals). We show the resulting black hole mass function. These constraints on the IMBH population have implications for the Laser Interferometer Space Antenna (LISA) mission and for cosmological models of black hole seeding and growth. We also constrain the extremely low luminosity end ($L_{\rm{bol}}\lesssim10^{40}$ erg s$^{-1}$) of the AGN luminosity functions at $z=0$. Our AGN luminosity functions are broadly consistent with an extrapolation of the shallow slope of the AGN luminosity functions from previous work.

\end{abstract}



\section{Introduction} \label{sec:intro}

Massive galaxies in the local Universe contain central supermassive black holes (SMBHs), but the population of intermediate-mass black holes (IMBHs) in dwarf galaxies remains poorly constrained \citep{Greene2020}. The key observable quantity is the so-called black hole occupation fraction (BHOF). The BHOF is the fraction of galaxies at a given stellar mass that contain a central, SMBH/IMBH \citep{Volonteri2008,Greene2012}. Despite their complex growth history via accretion and mergers, at $z=0$, the BHOF is expected to provide a fingerprint of SMBH seeding and growth, providing important constraints on theoretical models of SMBH evolution \citep{Volonteri2009,Lippai2009,Bellovary2011,Natarajan2011,Ricarte2018}.

To explain the formation of SMBHs at high redshifts when the Universe was only a few hundred Myr old \citep{Fragione2023,Andika2024}, SMBHs must have grown from early seed black holes (e.g., \citealt{Natarajan2014,Inayoshi2020}). Theories of SMBH seed formation channels broadly fall into two broad classes: ``light'' and ``heavy'' seeds. In the most popular light seeding scenario, black holes with masses of $\sim 10^{1-2}\ M_{\odot}$ are expected to form as remnants of the massive, first generation of stars, namely the Population III (Pop III) stars  \citep{Bond1984,Madau2001,Fryer2001,Abel2002,Bromm2003}. With improvement in the resolution of simulations that track the formation of first stars, it is now found that rather than forming individual stars, early star formation results in star clusters, whose evolution could also provide sites for the formation of light initial seeds \citep{Gurkan2004,PortegiesZwart2004}. Essentially, the light seed scenarios refer to starting with low mass seeds and gradually growing them over time. On the other hand, in the most popular ``heavy'' seeding scenario, black holes with masses of $\sim 10^4 - 10^6\ M_{\odot}$ are expected to viably form from direct collapse of pre-galactic disks under specific conditions \citep{Haehnelt1993,Loeb1994,Bromm2003,Koushiappas2004,Lodato2006,Begelman2006,Volonteri2008}. Additionally, multiple other formation channels have also been proposed, such as mechanisms that operate within nuclear star clusters \citep{Devecchi2009,Davies2011,Devecchi2010,Tal2014,Lupi2014,Antonini2015,Stone2017,Fragione2020,Kroupa2020,Natarajan2021}; inside globular clusters \citep{Miller2002,Leigh2014,Antonini2019}; and even young star clusters \citep{Rizzuto2021}. Heavy seeds are predicted to be fewer in number, while light seeds are predicted to be more abundant but less massive \citep{Volonteri2008}.

Observational limits on the low-redshift BHOF have been reported for galaxies with AGN-like optical emission line ratios \citep{Trump2015}, H$\alpha$ broad-line detections \citep{Cho2024}, dynamical black hole masses \citep{Nguyen2019}, using the nuclear star cluster occupation fraction as a proxy for the BHOF \citep{SJ2019,Hoyer2021}, and from local X-ray detections \citep{Miller2015}. \citet{Haidar2022} compared these various occupation fraction measurements to results from cosmological numerical simulations, finding significant discrepancies between them. This likely highlights the lack of consensus models that self-consistently include SMBH seeding, accretion modes, and especially AGN feedback. Meanwhile, observational measurements of the BHOF are limited by detectability and various selection biases.

Nuclear X-ray emission is a ubiquitous consequence of AGN accretion, and one of its most direct probes. \citet{She2017} identified 314 AGN candidates with distances less than 50 Mpc from the Chandra data archive. They estimate a BHOF of at least $\sim 21\%$ in local late-type galaxies. \citet{Pandya2016} searched the Chandra data archive for weakly-accreting central black holes in ultra compact dwarfs. Those sources are largely consistent with XRBs. Using \emph{Chandra} X-ray data from the volume-limited AGN Multiwavelength Survey of Early-Type Galaxies (AMUSE; \citealt{Gallo2010,Miller2012}), \citet{Miller2015} developed a Bayesian approach to infer the BHOF taking into account detectability limits. \citet{Gallo2019} improved upon these constraints and derived a resulting black hole mass function based on them. \citet{Birchall2022} identified a sample of X-ray $z<0.33$ AGN from the XMM-Newton serendipitous survey catalog and studied their accretion rate distributions, finding a constant active galaxy fraction with stellar mass that increases with redshift. That work was preceded by a similar analysis with SDSS \citep{Birchall2020}. \citet{Pardo2016} performed a similar analysis at $z \lesssim 1$, but only 10 sources were unambiguously identified as AGNs.

Like X-ray emission, possibly all massive galaxies have compact radio emission originating from an AGN when sensitivity permits its detection. The radio core luminosity scales with host galaxy stellar mass, with the fraction of nearby galaxies with radio core emission approaching unity at high stellar masses \citep{Capetti2009,Baldi2018,Ito2022}. For radio-loud galaxies, the radio emission is thought originate from synchrotron emission from an unresolved base of a jet. For radio quiet galaxies, the origin of the observed radio core emission is more controversial (e.g., \citealt{Balmaverde2006,Blundell2007,Laor2008,Panessa2019}).

Complementary to X-ray and radio, searches for optical variability in low mass galaxies in wide-field surveys is becoming an established technique for identifying AGN by probing accretion disk flux variations \citep{Baldassare2018,Baldassare2020,Ward2022}. The amplitude of flux variations are known to scale with the black hole mass and AGN luminosity \citep{Kelly2009,MacLeod2010}. However, optical variability can only detect optically-unobscured AGNs that are luminous enough to overcome the optical emission from the host galaxy continuum. This combination of effects reduces the detected fraction of optical variability in nearby galaxies to $1-2$ percent \citep{Burke2023}.
 
In this paper, we infer the intrinsic BHOF in local galaxies from detections of nuclear core activity. We accomplish this by constructing a censored hierarchical Bayesian generalized linear model \citep{deSouza2015} that describes the fractional X-ray, radio, and optical variability activity that scales with host galaxy stellar mass. We introduce our model in \S\ref{sec:methods} and validate it using mock data generated from a simple forward model. The data is described in \S\ref{sec:data}. In \S\ref{sec:result}, we present our inferred BHOF results from published radio, X-ray, and optical variability catalogs. In \S\ref{sec:bhmf}, we use the $z=0$ galaxy stellar mass function (GSMF; \citealt{Driver2022}) and the black hole -- host galaxy stellar mass relations ($M_{\rm{BH}}-M_{\star}$; \citealt{Reines2015,Greene2020}) to obtain multi-wavelength constraints on the black hole mass function below $M_{\rm{BH}} \sim 10^6 M_{\odot}$. In \S\ref{sec:discussion}, we discuss implications for the Laser Interferometer Space Antenna (LISA) mission and compare to BHOF constraints from other probes. We conclude and discuss future directions in \S\ref{sec:conclusions}.

\section{Method to infer the occupation fraction} \label{sec:methods}

The BHOF is the fraction of galaxies at a given stellar mass that contain a central black hole more massive than a stellar-mass black hole. The simplest way an observer might try to measure the BHOF is by plotting the fraction of galaxies that have a detectable black hole signature versus host galaxy stellar mass. A limitation of this approach is that the AGN luminosity is expected to scale with the mass of the host galaxy, assuming a mass-independent Eddington ratio distribution and radiative efficiency, and a black hole mass that scales with the host galaxy stellar mass with a fixed scatter. This means that lower-mass galaxies would have fewer detections as a larger fraction of the source fluxes scatter below the detection limit. Disentangling the intrinsic BHOF from detectability is difficult for dwarf galaxies containing IMBHs with luminosities close to the detection limit. 

The black hole mass function can be inferred directly if the details of the selection function are known (e.g., \citealt{Kelly2013,Weigel2016,Ananna2022,Cho2024}). In this section, we develop a Bayesian inference formalism to robustly infer the intrinsic BHOF from cataloged data of AGN accretion signatures. Our approach uses detections and non-detections (upper-limits) to infer the BHOF without the need to directly model the selection function. We begin by considering the case of X-ray searches for AGN activity in local galaxies as in \citet{Miller2015}. Later, we will generalize the approach to two other probes of AGN activity: radio and optical variability. 

\subsection{Description of the Bayesian inference approach}

\begin{figure}
\centering
\includegraphics[width=0.45\textwidth]{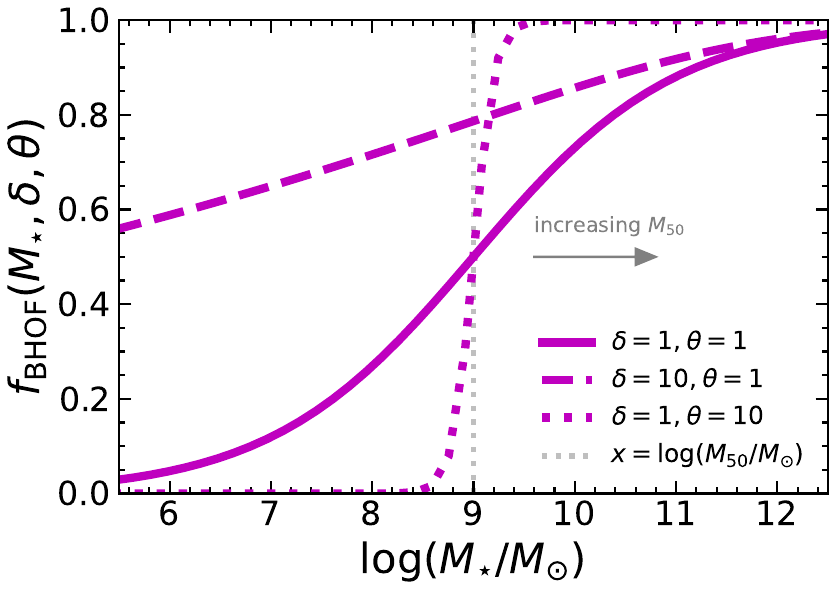}
\caption{Functional form of the black hole occupation fraction ($f_{\rm{BHOF}}$) as function of total galaxy stellar mass for $\log(M_{\rm{50}}/M_{\odot})=9$ (marked by the vertical dotted line). The parameter $\delta$ controls the slope near the inflection point and $\theta$ is the growth rate from zero to full occupation and $\delta$ fixes the inflection point. Increasing $M_{\rm{50}}$ shifts the inflection point to the right, lowering the occupation fraction at low stellar masses.
\label{fig:fbhof}}
\end{figure}

\begin{figure*}
\centering
\includegraphics[width=0.45\textwidth]{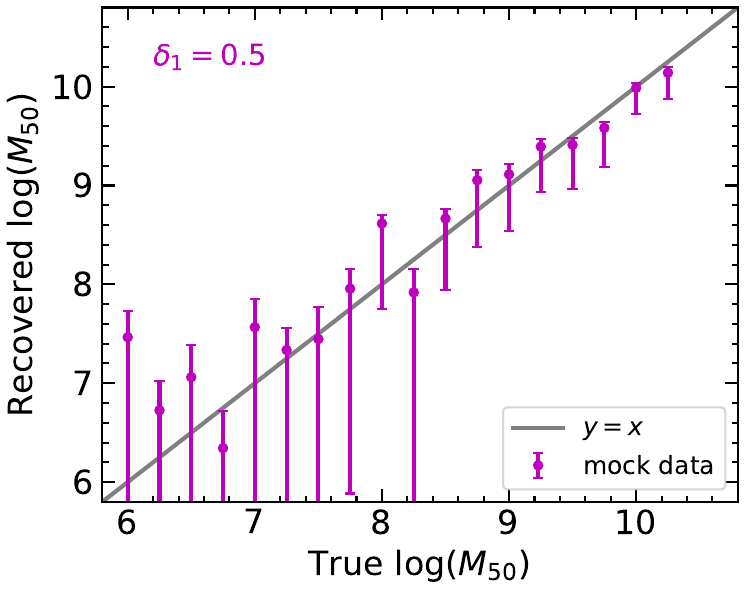}
\includegraphics[width=0.45\textwidth]{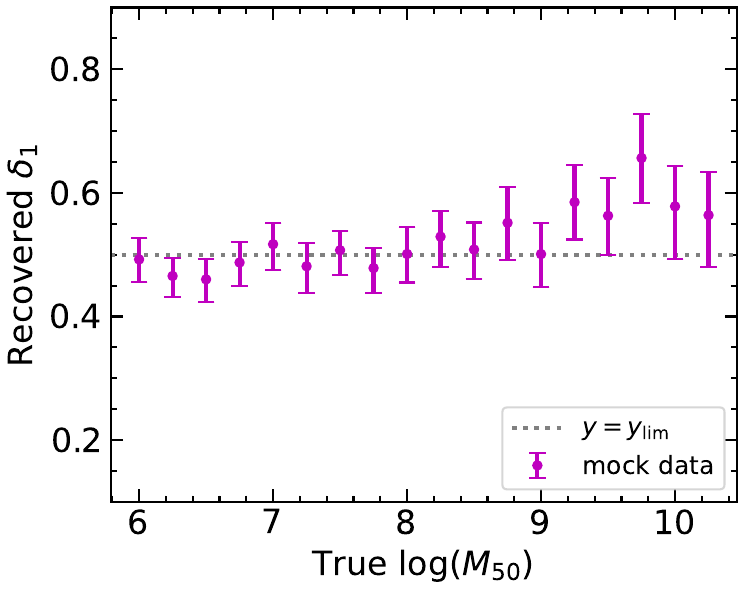}
\caption{Recovered BHOF (\emph{left}) and ERDF power-law slope $\delta_1$ (\emph{right}). The left panel is the recovered versus true inflection point in the black hole occupation function $M_{50}$ using mock data generated following the procedure described in the text. A larger $M_{50}$ value corresponds to a \emph{lower} BHOF at low stellar masses. The uncertainties on $M_{50}$ are given by the 5th and 95th percentiles of the posterior distributions. The uncertainties on $\beta$ are given by the 16th and 84th percentiles ($\approx 1\sigma$) of the posterior distributions. The corresponding occupation fractions at low $M_{50}$ are lower limits on the BHOF, because the data are deep enough to constrain inflection point in the occupation function. The inferred parameters are generally consistent with the true values to $\sim {1}\sigma$. The uncertainty on the recovered ERDF increases as $M_{50}$ increases, because there are fewer detected data points to constrain the power-law slope $\delta_1$. 
\label{fig:mock}}
\end{figure*}

We assume a log-linear relation between the AGN luminosity in a given band $L$ and the stellar mass $M_{\star}$, with measurement errors $\sigma_L$ and $\sigma_{M_{\star}}$ and scatter in luminosity at fixed mass given by the accretion rate distribution. These two observed variables are related through the definition of the Eddington ratio as,
\begin{multline}
        \label{eq:lambdaEdd}
    \log \lambda = \log(L_{\rm bol} / {\rm erg\ s}^{-1}) + \log {\rm{BC}} -38.2 \\ -\log(M_{\rm{BH}}/M_{\odot}),
\end{multline}
where $L_{\rm{bol}}$ is the AGN bolometric luminosity, $M_{\rm{BH}}$ is the black hole mass, and BC is the bolometric correction, $\log L_{\rm{bol}} = \log L + \log {\rm{BC}}$. We use the following bolometric corrections:
\begin{equation}
\log {\rm{BC}} = 
\begin{cases} 
      1 & {\text{X{-}ray}} \\
      3 & {\rm radio} \\
      0.7 & {\rm optical\ var}.
   \end{cases}
\end{equation}
The X-ray and radio bolometric corrections are identical to those used in \citet{Weigel2017}. Our adopted optical BC is the approximate average from \citet{Duras2020} and \citet{Marconi2004}. In lieu of actual BH mass measurements, we assume a black hole mass -- stellar mass relation,
\begin{equation}
    \label{eq:mbhmstar}
    \log(M_{\rm{BH}}/M_{\odot}) = \alpha_\star + \beta_\star\ \big(\log(M_{\star}/M_{\odot}) - 10.5\big),
\end{equation}
where the we adopt the coefficients from \citet{Greene2020} that depend on galaxy type:
\begin{equation}
\alpha_\star = 
\begin{cases} 
      \mathcal{N}(6.70, 0.13) & {\text{X{-}ray}} \\
      \mathcal{N}(7.89, 0.09) & {\rm radio} \\
      \mathcal{N}(6.70, 0.13) & {\rm optical\ var},
   \end{cases}
\end{equation}
\begin{equation}
\beta_\star = 
\begin{cases} 
      \mathcal{N}(1.61, 0.24) & {\text{X{-}ray}} \\
      \mathcal{N}(1.33, 0.12) & {\rm radio} \\
      \mathcal{N}(1.61, 0.24) & {\rm optical\ var},
   \end{cases}
\end{equation}
assuming the X-ray and optical variability samples select predominately blue late-type galaxies and the radio sample selects mainly early-type galaxies with more massive black hole to stellar mass ratios \citep{Weigel2017}. These fitting coefficients are taken directly from \citet{Greene2020}. The black hole mass -- stellar mass relation for optical AGNs is similar to that of inactive late-type galaxies \citep{Reines2015}.

Different normalizations of the black hole mass -- stellar mass relation will simply change the mean Eddington ratio of the sample without affecting the inferred black hole occupation fraction. At fixed stellar mass, the observed AGN luminosity probability distribution is given by the specific Eddington ratio distribution function (sERDF) and the scatter in the black hole mass -- stellar mass relation. The measurement uncertainties and the intrinsic scatter in this relation and in the bolometric corrections will be folded into the ERDF by convolving the intrinsic Eddington ratio distribution function (ERDF) with a normal distribution.

The observed X-ray luminosity probability distribution is a mixture of three components. The first component is the product of the sERDF with the probability that the black hole can be detected. The second component is probability that the black hole cannot be detected (due to it being  quiescent-state or inactive, obscured, or non-existent) and a distribution centered near $L=0$ (Bernoulli process). The third component is the probability that the luminosity of the source originates from star-formation-related contamination, rather than AGN emission. The total probability distribution is,
\begin{multline}
p(\log L | \log M_{\star}) = 
p(I | \log M_{\star})\ \bigg[ (1 - P_{\rm{SF}})\ \times \\
\bigg( f_{\rm{d}}\ {\rm{sERDF}}(\lambda | \log M_{\star}) +
(1-f_{\rm{d}})\ \delta(\log L + 9999) \bigg) \\ + P_{\rm{SF}}\ \mathcal{N}(\log L - \log L_{\rm{SF}} | s^2_{\rm{SF}}) \bigg],
\label{eq:prob}
\end{multline}
where ${\rm{sERDF}}({\lambda}| \log M_{\star})$ is the specific Eddington ratio distribution function (the ERDF at fixed stellar mass), $\delta$ is the Dirac $\delta$ function, $f_{\rm{d}}$ is the fraction of black holes that are detectable, $P_{\rm{SF}}$ is the probability that the luminosity is due to non-AGN emission $\log L_{\rm{SF}}$, such as an XRB.\footnote{Our black hole vs. star-formation origin mixture model is analogous to the foreground vs. background model of \citet{Hogg2010}.} The term in parenthesis is analogous to a Bernoulli probability distribution with probability $f_{\rm{d}}$. In the X-ray, the model XRB luminosity $\log L_{\rm{SF}}$ scales with the galaxy stellar mass and star formation rate (SFR) \citep{Lehmer2010}. The background quasar or cosmic X-ray background contributions are negligible within a small aperture centered on the galaxy nucleus.

The $p(I{=}1)$ term is the probability that a galaxy from the total astrophysical population is included in our parent sample, whether or not it is detected in the X-ray. The latent variable $I_i=1$ represents a source in the observed sample, while $I_i=0$ represents an unobserved source that is not present in the parent galaxy catalog due to incompleteness. We assume that the sample selection is independent of AGN luminosity, because we have included non-detections in our model. Under these conditions, the inference is independent of selection effects in the parent galaxy catalog \citep{Kelly2007}. We will determine the completeness $p(I{=}1 | \log M_{\star})$ empirically when inferring on real data. 

We assume a mass-independent smoothly broken power law ERDF (e.g., \citealt{Aird2012,Aird2013,Caplar2015,Weigel2017,Carraro2022}). The probability distribution of our ERDF is,
\begin{equation}
{\rm{ERDF}}(\lambda) = C \left[ \left(\frac{\lambda}{\lambda_\ast}\right)^{\delta_1} + \left(\frac{\lambda}{\lambda_\ast}\right)^{\delta_1+\epsilon} \right]^{-1}
\end{equation}
where $\delta_1$ is the power-law index of the low-$\lambda$ end of the intrinsic ERDF, $\delta_1+\epsilon$ is the power-law index of the high-$\lambda$ end of the intrinsic ERDF, $\lambda_\ast$ is the transition Eddington ratio between power laws, and $C$ is a normalization constant where $\lambda_{\rm min}$. A smoothly-broken power law distribution has been demonstrated to be a better description of the intrinsic ERDF of low-redshift AGNs compared to a single power law, Schechter, or lognormal distribution over a wide range of luminosities (e.g., \citealt{Aird2013,Weigel2017}). The high end slope $\delta_2$ is set by the high luminosity slope of the function. The low end slope $\delta_1$ is set by either the low luminosity slope of the luminosity function or the low-mass slope of the black hole mass function, whichever is greater. The transition Eddington ratio $\lambda_\ast$ is roughly set by the break in the luminosity function.

The probability distribution function of the sERDF is given by convolving the ERDF with a normal distribution,
\begin{multline}
    \label{eq:erdf}
    {\rm{sERDF}}(\lambda) = \\ \int^{\lambda_{\rm max}}_{\lambda_{\rm{min}}} {\rm ERDF}(\lambda^\prime)\ \mathcal{N}(\log \lambda - \log \lambda^{\prime}| 0, s^2)\ d \lambda^{\prime},
\end{multline}
\begin{equation}
C^{-1} = \int_{\lambda_{\rm min}}^{\lambda_{\rm max}} {\rm sERDF}(\lambda)\ d\lambda,
\end{equation}
where $\lambda_{\rm{min}}$ and $\lambda_{\rm{max}}$ is the minimum and maximum Eddington ratio, $\mathcal{N}(x | \mu, s^2)$ is a normal distribution with mean $\mu$ and variance $s^2 = (\beta_\star\ \sigma_{\log M_{\star}})^2 + (\sigma_{\log L})^2 + \sigma^2$, where $\sigma$ is the intrinsic scatter.\footnote{In this treatment, we assume no covariance between the error terms. More general approaches exist \citep{Kelly2007,Hogg2010}.}

We model the \emph{detectable} fraction as the product of the occupation fraction, the active fraction, and the fraction of sources that are unobscured along our line of sight as,
\begin{equation}
    f_{\rm{d}} = f_{\rm{BHOF}} \times f_{\rm{a}} \times (1 - f_{{\rm obsc}}),
\end{equation}
where $f_{\rm{a}}$ is the active fraction, $f_{\rm{BHOF}}$ is the BHOF, and $f_{\rm obsc}$ is the obscured fraction in the band. We model the BHOF using a generalized logistic function \citep{Richards1959}:
\begin{equation}
        f_{\rm{BHOF}}(\log M_{\star}) = \frac{1}{[1 + \delta \exp (-\theta (\log M_{\star} - \log M_{50}))]^{1/\delta}}
\label{eq:BHOF}
\end{equation}
where $\log M_{50}$ is the stellar mass inflection point in the logistic function, $\theta$ is the growth rate from zero to full occupation, and $\delta$ fixes the infection point (Figure~\ref{fig:fbhof}). This simple functional form is chosen because it is bound between 0 and 1, and it flexibly but smoothly models the possibilities between both heavy and light seeding scenarios. Our Equation~\ref{eq:BHOF} is analogous to, but more general than, the BHOF used in \citet{Miller2015}. The impact of different choices of the functional form for the BHOF is described in Appendix~\ref{sec:amuse}. We eliminate the degeneracy between $\delta_1$ and the low mass slope of the black hole mass function by assuming a flexible functional form for the BHOF but a mass-independent ERDF.

The detectable fraction will be defined depending on the observed wavelength. For the X-ray, we will fix $f_{\rm{obsc}} = 0$, because we are observing in the hard band. Any diminishing of the observed luminosity due to heavy obscuration in the hard X-ray band will be folded into the active fraction and the scatter in the sERDF. We will also assume no obscuration in the radio. However, the optically obscured fraction depends on the incidence of an obscuring torus at a given AGN luminosity (e.g., \citealt{Urry1995}). For the optical variability data, we adopt the functional form of the Type 1 AGN fraction derived from a modified receding torus model \citep{Simpson2005,Oh2015}:
\begin{equation}
f_{{\rm obsc}} = 
\begin{cases} 
      0 & {\text{X{-}ray}} \\
      0 & {\rm radio} \\
      [1 + 3(L_{\star}/L_0)^{1-2\xi}]^{-a} & {\rm optical\ var} 
   \end{cases}
\end{equation}
where $L_0$ is a pivot luminosity and $a$ and $\xi$ govern the shape of the function. The shape parameters $a$ and $\xi$ are difficult to constrain with our data and somewhat degenerate with the active fraction. Therefore, we adopt normally-distributed priors of $a \sim \mathcal{N}(0.52, 0.05)$ and $\xi \sim \mathcal{N}(0.23, 0.02)$, taking the fitted values from \citet{Oh2015}.


We take into account contamination from XRBs in the X-ray and from synchrotron emission from H II regions in the radio. The contribution from star formation related emission is:
\begin{equation}
\log L_{{\rm SF}} = 
\begin{cases} 
      \log L_{\rm{XRB}} + {\rm CI} & {\text{X{-}ray}} \\
      \log S_{24~\mu\rm{m}} + {\rm CI} - q_{\rm{TIR}} & {\rm radio} \\
      -9999 & {\rm optical\ var}
   \end{cases}
\end{equation}
where $L_{\rm{XRB}}$ galaxy-wide XRB luminosity \citep{Lehmer2010}, and $\log S_{24~\mu\rm{m}}$ is the expected radio luminosity of the galaxy from star formation related emission. For the radio data, $q_{\rm{TIR}}$ is a ratio of the expected radio luminosity from star formation processes given the galaxy infrared luminosity (e.g., \citealt{Condon1992}). We assume no contamination from variable star formation related processes for the optical variability data, because obvious supernova and other transients have been removed from the data we will use.

When matching catalogs with a radius centered on the galaxy center, we estimate the star formation or XRB contribution within the search radius by assuming the stellar population density traces the light profile of the galaxy. We assume a galaxy type dependent Sérsic light profile of the form $I(r) \propto \exp(-b_n\ r^{1/n})$ and calculate the fraction of light within the nucleus using the concentration index (CI) \citep{Trujillo2001} as:
\begin{equation} \label{eq:nucleation}
    {\rm CI} = \log \gamma \big(2n, b_n (2 r_s/d_{25})^{1/n}\big) - \log \gamma(2n, b_n),
\end{equation}
where $d_{25}$ is the galaxy diameter, $r_s$ is the cross matching (search) radius, $n$ is the Sérsic index, $b_n \approx 2n - 1/3$, and $\gamma$ is the incomplete gamma function. We will crudely assume $n=4$ for early-type galaxies (de Vaucouleurs's profile) and $n=1$ for late-type galaxies. This component of the model is necessary because if the XRB population is over-estimated, the lower luminosity points will be down-weighted in the likelihood. 

The standard likelihood including both detections, non-detections with upper limits is \citep{Kelly2007,Isobe1986}:
\begin{multline}
    \mathcal{L} = \prod_{i \in \mathcal{A}_{\rm det}} p(\log L_i | \log M_{{\star},i})\ \times \\
    \prod_{j \in \mathcal{A}_{\rm cens}} \int^{\log L_j}_{-\infty} p(\log L_j | \log M_{{\star},j})\ d \log L_j.
\end{multline} 
The first term is the product over the detected data set. The second term is the product over the censored data set containing the non-detections with upper limits $\log L_j$. The second term is marginalized over all possible values of non-detections. The resulting log likelihood is given in Appendix~\ref{sec:logL}. The radio and optical variability likelihood function is identical to that above without the XRB contribution term ($P_{\rm{XRB}} = 0$).

The expected observed detection fraction given a simple luminosity-limited sample is:
\begin{equation}
\label{eq:exp}
    \big\langle f_{\rm{d}}(\log M_{\star}) \big\rangle = f_{\rm{d}}(\log M_{\star}) \frac{ \int_{\lambda_{\rm{lim}}}^{\lambda_{\rm{max}}} {\rm{sERDF}}(\lambda)\ d\lambda }{ \int_{\lambda_{\rm{min}}}^{\lambda_{\rm{max}}} {\rm{sERDF}}(\lambda)\ d\lambda }.
\end{equation}
where $\log \lambda_{\rm{lim}} = \log(L_{\rm{lim}} / {\rm{erg\ s}}^{-1}) + \log {\rm{BC}} -38.2 - \log(M_{\rm{BH}}/M_{\odot})$, where $L_{\rm{lim}}$ is the luminosity detection limit.

Given the limited dynamic range of luminosities at low redshift, we assume priors on the ERDF parameters from the fitting from \citet{Weigel2017}. We assume uninformative and uniform priors on the BHOF parameters. Our priors are given in Table~\ref{tab:priors}. We use standard Markov chain Monte Carlo techniques via the \textsc{emcee} code \citep{Foreman-Mackey2013} to sample from the posterior probability density. We use a mixture of differential evolution \citep{Nelson2014} and snooker proposals \citep{terBraa2008}, which we found converged faster than the default stretch move proposal \citep{Goodman2010}. We then sample from the posterior predictive distribution of the BHOF. We estimate the 1$\sigma$ uncertainties on the BHOF by taking the 16th and 84th percentiles of the posterior predictive distribution of the BHOF. Therefore, we are able to jointly constrain the ERDF and the BHOF by relating these astrophysical quantities to the observed ones.

\begin{deluxetable}{ll}[t!]
\label{tab:priors}
\tablewidth{0.43\textwidth}
\tablecaption{Summary of model parameters and prior distributions}
\tablehead{
\colhead{Parameter} & \colhead{Prior Distributions}
}
\startdata
$\log(M_{\star, 50} / M_{\odot})$  & $\mathcal{U}(5, 12)$   \\
$\delta$  & $\mathcal{U}(0, 50)$   \\
$\theta$  & $\mathcal{U}(0, 50)$   \\
$\delta_{1,\ 2-10\ {\rm keV}}$  & $\mathcal{N}(0.47, 0.31)$\tablenotemark{a}   \\
$\epsilon_{2-10\ {\rm keV}}$  & $\mathcal{N}(2.22, 0.41)$\tablenotemark{a}   \\
$\log \lambda_{\ast,\ {2-10\ {\rm keV}}}$  & $\mathcal{N}(-1.84, 0.34)$\tablenotemark{a}   \\
$0.5 \ln(\sigma_{2-10\ {\rm keV}})$  & $\mathcal{U}(-2, 1)$   \\
$f_{a,\ 2-10\ {\rm keV}}$  & $\mathcal{U}(0, 1)$   \\
$P_{\rm SF,\ 2-10\ {\rm keV}}$  & $\mathcal{U}(0, 1)$ \\
$\delta_{1,\ 3\ {\rm GHz}}$  & $\mathcal{N}(0.41, 0.02)$\tablenotemark{a}   \\
$\epsilon_{3\ {\rm GHz}}$  & $\mathcal{N}(0.82, 0.16)$\tablenotemark{a}   \\
$\log \lambda_{\ast,\ 3\ {\rm GHz}}$  & $\mathcal{N}(-2.81, 0.18)$\tablenotemark{a}   \\
$0.5 \ln(\sigma_{3\ {\rm GHz}})$  & $\mathcal{U}(-2, 1)$   \\
$f_{a,\ 3\ {\rm GHz}}$  & $\mathcal{U}(0, 1)$   \\
$P_{\rm SF,\ 3\ {\rm GHz}}$  & $\mathcal{U}(0, 1)$ \\
$\delta_{1,\ R}$  & $\mathcal{N}(0.47, 0.31)$\tablenotemark{a}   \\
$\epsilon_{R}$  & $\mathcal{N}(2.22, 0.41)$\tablenotemark{a}   \\
$\log \lambda_{R}$  & $\mathcal{N}(-1.84, 0.34)$\tablenotemark{a}   \\
$0.5 \ln(\sigma_{R})$  & $\mathcal{U}(-2, 1)$   \\
$f_{a,\ R}$  & $\mathcal{U}(0, 1)$ \\
$P_{{\rm SF},\ R}$  & $=0$ \\
$\log(L_0/ {\rm erg\ s}^{-1})$  & $\mathcal{U}(40, 50)$ \\
$\xi$  & $\mathcal{N}(0.23, 0.02)$\tablenotemark{c} \\
$a$  & $\mathcal{N}(0.52, 0.05)$\tablenotemark{c}
\enddata
\tablecomments{$\mathcal{U}(x, y)$ denotes a uniform distribution between $(x,y)$. $\mathcal{N}(\mu, s^2)$ denotes a normal distribution with mean $\mu$ and variance $s$.}
\tablenotetext{a}{Priors taken from \citet{Weigel2017} after symmetrizing the parameter uncertainties by taking the average. }
\tablenotetext{c}{Priors taken from \citet{Oh2015}.}
\end{deluxetable}

\subsection{Validation of the Bayesian inference approach}

To validate our Bayesian inference pipeline, we generate mock data with a known BHOF and ERDF. We will also assign upper limits to AGN luminosities below a chosen luminosity limit to mimic a simple luminosity-limited sample. Then, we will use the likelihood function to infer the BHOF. We will repeat this procedure, varying the true BHOF. A successful validation will result in an inferred BHOF consistent with the known, true BHOF. The result is shown in Figure~\ref{fig:mock} and detailed below. 

We vary the pivot mass of the BHOF between $\log(M_{50} / M_{\odot}) = 5-12$ and draw random samples from the ERDF with $\delta_1=0.5$, $\epsilon=2.0$, and a bolometric correction of $10$ to roughly match the previous work from X-ray data. We set the AGN luminosity of unoccupied galaxies to zero after drawing from a Bernoulli distribution with the occupation probability given by Equation~\ref{eq:BHOF}. Finally, we apply a luminosity limit of $10^{38}$ erg s$^{-1}$, such that sources with true AGN luminosity below this value are assigned an upper-limit of $10^{38}$ erg s$^{-1}$. Different measurement uncertainties, luminosity limits, and ERDF parameters will affect the resulting BHOF limits; we have chosen reasonable values to illustrate the capability of our inference pipeline. Our inferred values are consistent with our mock data, as shown in Figure~\ref{fig:mock}.

\section{Data} \label{sec:data}

\subsection{X-ray data}

In order to obtain X-ray fluxes and upper-limits for local galaxies, we matched publicly-available X-ray catalogs to the 50 Mpc Galaxy Catalog\footnote{\url{https://github.com/davidohlson/50MGC}} (50MGC; \citealt{Ohlson2024}). The 50MGC catalog is a catalog of 15,424 galaxies within 50 Mpc with homogenized stellar masses estimated from their mass-to-light ratios. The 50MGC catalog combines sources from HyperLeda \citep{Makarov2014}, the NASA-Sloan Atlas \citep{Blanton2011}, and the Catalog of Local Volume Galaxies \citep{Karachentsev2013}.

We identified a number of higher redshift galaxies with incorrect spectroscopic redshifts in the 50MGC catalog, mostly originating from the NASA-Sloan Atlas. The NASA-Sloan Atlas redshifts are from SDSS data release 11. Therefore, we use the updated spectroscopic redshifts from SDSS data release 16 \citep{SDSSDR162020} and GAMA data release 3 \citep{Baldry2018} wherever available. We also removed sources near the Fornax (within 1$^{\circ}$ of $54.6225^{\circ}$, $-35.4522^{\circ}$) and Virgo (within 12$^{\circ}$ of $186.6338^{\circ}$, $12.7233^{\circ}$) clusters. The BHOF in galaxy clusters is expected to differ substantially from the predictions for local field galaxies \citep{Tremmel2023}. In the sections below, we detail the X-ray catalogs that we have used.

\subsubsection{Chandra Source Catalog}

We matched version 2.1 of the Chandra Source Catalog \citep{Evans2024} to our parent sample of galaxies from the 50 Mpc Galaxy Catalog\footnote{\url{https://github.com/davidohlson/50MGC}} (50MGC; \citealt{Ohlson2024}) using a 0.5$^{\prime\prime}$ search radius. This version of the Chandra Source Catalog includes data from observations taken before the end of 2014. We use the coadded aperture fluxes in fields with multiple Chandra observations.

We use the \texttt{flux\_aper\_h} values from the the Chandra Source Catalog using the \textsc{CSCview} software\footnote{\url{https://cxc.cfa.harvard.edu/csc/}} to obtain the $2{-}7$ keV coadded fluxes. The conversion from photons to flux takes into account the response function, PSF, and exposure time without assuming any spectral model. The $2{-}7$ keV band minimizes absorption effects in the soft X-ray bands from Compton thick AGNs. Then, we restricted our dataset to compact X-ray sources only using the \texttt{extent\_flag} parameter. Extended X-ray detections are treated as upper limits on the compact AGN emission for the Chandra data and the additional data described below. We obtained flux upper limits for sources with Chandra observations but without detections from the \texttt{flux\_sens\_h} parameter. See \citet{Evans2010,Evans2024} for details of the flux and source detections calculations.

We convert the X-ray flux to luminosity using, 
\begin{equation} \label{eq:LX}
L_{2-10\ {\rm keV}} = 4 \pi d^2\ \frac{10 ^{2-\Gamma} - 2^{2-\Gamma}}{E_2^{2-\Gamma} - E_1^{2-\Gamma}}\ f_{E_2-E_1},
\end{equation}
where $L_{2{-}10\ {\rm keV}}$ is the $2{-}10$ keV luminosity, $d$ is the distance given in the 50MGC catalog, $\Gamma$ is the spectral index, $f_{E_2-E_1}$ is the flux from the Chandra Source Catalog, and $E_2$ and $E_1$ are the energies in keV from the flux band which we are converting. We assume $\Gamma=1.8$ throughout this work, which is typical for low-luminosity AGNs \citep{Ho2008,Ho2009}. The conversion from $2{-}7$ keV to $2{-}10$ keV is done to consistently compare against the $2{-}10$ keV XRB luminosity scaling relations in the literature, but does not significantly impact our results. We have ignored the $K$-correction for a given X-ray spectral index, which is small at these distances of $< 50$ Mpc. 

\subsubsection{XMM-Newton Serendipitous Source Catalog}

We supplement the Chandra data by matching our parent sample of galaxies to the fourth XMM-Newton Serendipitous Source Catalog \citep{Webb2020} using a $1^{\prime\prime}$ search radius. We restrict the catalog to point sources using the \texttt{ext} column and with no data quality flags. We use the 0.2-12 keV \texttt{flux8} values, and convert them to the 2-10 keV band using Equation~\ref{eq:LX}. Upper limits from non-detections from XMM-Newton observations are not easily obtainable for our entire parent sample. Therefore, we will adopt upper limits from shallower all-sky X-ray surveys ROSAT and eROSITA. The XMM-Newton upper limits are generally consistent with, and much more stringent than, the ROSAT or eROSITA all sky X-ray surveys so this assumption will not affect our proceeding analysis.

\subsubsection{eROSITA}

We matched our parent sample of galaxies to the data release 1 eROSITA German sky portion of the all sky eROSITA survey \citep{Merloni2024} using a $1.5^{\prime\prime}$ search radius. We restrict the catalog to point sources using the \texttt{EXT} column. We use the hard band 2.3-5 keV flux values, and convert them to the 2-10 keV band using Equation~\ref{eq:LX}. The eROSITA data release 1 flux limit of $\sim 6.5 \times 10^{-15}$ erg cm$^{-2}$ s$^{-1}$ \citep{Tubin-Arenas2024} corresponds to a luminosity limit of $\sim 1.9\times 10^{39}$ erg s$^{-1}$ at 50 Mpc, which is much shallower than the Chandra observations with a typical luminosity limit of $\sim 10^{38}$ erg s$^{-1}$. We adopt conservative upper limits of $10^{-14}$ erg cm$^{-2}$ s$^{-1}$ for sources without a Chandra, XMM-Newton, or eROSITA detection that lie in the Western Galactic hemisphere $359.9^{\circ} > l > 179.9^{\circ}$.

\subsubsection{ROSAT}

Finally, we matched our parent sample of galaxies to the second ROSAT all-sky survey point source catalog \citep{Boller2016} using a $10^{\prime\prime}$ search radius. The astrometric precision of $\sim 20^{\prime\prime}$ is far too large to confidently associate the detections with our galaxy centers. Therefore, ROSAT matches to sources without a detection from any of the previous catalogs are assumed to be upper limits on the compact AGN emission. These conservatively-chosen upper limits would still be consistent with cases where the detections and associations are genuine. Sources without a Chandra, XMM-Newton, eROSITA, or ROSAT detection are assumed to have an upper limit of $10^{-13}$ erg cm$^{-2}$ s$^{-1}$. We convert the 0.1-2.4 keV flux values to the 2-10 keV band as usual using Equation~\ref{eq:LX}.

\begin{deluxetable*}{lcccccccccccc}[t!]
\label{tab:xray}
\tablewidth{0.98\textwidth}
\tablecaption{Galaxies with compact X-ray detections}
\tablehead{
\colhead{\texttt{objname}} & \colhead{\texttt{ra}} & \colhead{\texttt{dec}} & \colhead{\texttt{type}} & \colhead{\texttt{dist}} & \colhead{\texttt{e\_dist}} & \colhead{\texttt{logmass}} & \colhead{\texttt{e\_logmass}} & \colhead{\texttt{logLX}} & \colhead{\texttt{e\_logLX}} & \colhead{\texttt{logSFR}} & \colhead{\texttt{e\_logSFR}}  & \colhead{\texttt{p\_{{SF}}}} \\
& \colhead{(deg)} & \colhead{(deg)} & & \colhead{(Mpc)} & \colhead{(Mpc)} & \colhead{($\log M_{\odot}$)} & \colhead{(dex)} & \colhead{($\log$ erg s$^{-1}$)} & \colhead{(dex)} & \colhead{($\log M_{\odot}$ yr$^{-1}$)} & \colhead{(dex)} & \colhead{}
}
\startdata
2MASXJ1515 & 228.754 & 55.432 & early & 49.675 & 6.09 & 9.663 & 0.136 & 39.02 & 0.2 & -1.74 & 0.3 & 0.0 \\
ESO103-056 & 280.8879 & -64.1068 & early & 39.289 & 5.909 & 9.82 & 0.136 & 39.41 & 0.2 & -0.27 & 0.3 & 0.0 \\
ESO121-006 & 91.8741 & -61.8076 & late & 14.142 & 8.713 & 9.472 & 0.624 & 40.24 & 0.38 & -1.0 & 0.3 & 0.0 \\
ESO320-030 & 178.2989 & -39.1302 & late & 38.38 & 9.589 & 10.405 & 0.222 & 40.0 & 0.16 & 0.81 & 0.3 & 0.0 \\
ESO383-035 & 203.9741 & -34.2956 & late & 26.885 & 5.093 & 9.75 & 0.167 & 42.44 & 0.12 & 0.18 & 0.3 & 0.0 \\
\dots & \dots & \dots & \dots & \dots & \dots & \dots & \dots & \dots & \dots & \dots & \dots
\enddata
\tablecomments{\texttt{objname} - Common galaxy name (\texttt{objname}); \texttt{ra} - Right ascension (\texttt{ra}); \texttt{dec} - Declination (\texttt{dec}); \texttt{type} - Galaxy type (\texttt{best\_type}); \texttt{dist} - Distance (\texttt{bestdist}); \texttt{e\_dist} - Error on the distance (\texttt{bestdist\_error}); \texttt{logmass} - Logarithm of the galaxy stellar mass (\texttt{logmass}); \texttt{e\_logmass} - Error on the logarithm of the galaxy stellar mass (\texttt{logmass\_error}); \texttt{logLX} - Logarithm of the adopted 2$-$10 keV X-ray luminosity; \texttt{e\_logLX} - Error on the logarithm of the adopted 2$-$10 keV X-ray luminosity (excluding contribution from distance error); \texttt{logSFR} - Logarithm of the estimated star formation rate; \texttt{e\_logSFR} - Adopted error on the logarithm of the estimated star formation rate (values of 0.3 dex indicate that the SFR is estimated from the stellar mass $-$ SFR relation of \citet{Dale2023}); \texttt{p\_SF} - Posterior probability that the luminosity is due to X-ray binary emission rather than black hole accretion. The names in parenthesis are the corresponding column names from the \citet{Ohlson2024} catalog. These column data are repeated here for completeness. Only a portion of the table is shown here. The full version is available as supplementary data.}
\end{deluxetable*}

\subsubsection{AGN Targeting Bias in Chandra and XMM-Newton Observations}

Presumably, a disproportionate fraction of Chandra and XMM-Newton observations in our dataset are targeted observations of nearby AGNs. However, we assumed that our observations were unbiased toward black hole occupation when deriving our likelihood function. Therefore, we need to include X-ray detections or upper limits for every source in our parent sample to perform unbiased inference. Therefore, we included upper limits from eROSTIA or ROSAT all-sky catalogs wherever applicable, as desribed above. Now every source in our parent sample has an X-ray detection or upper limit and we can proceed with an unbiased inference of the BHOF.

\subsubsection{Completeness of parent sample}

\begin{figure}[ht!]
\plotone{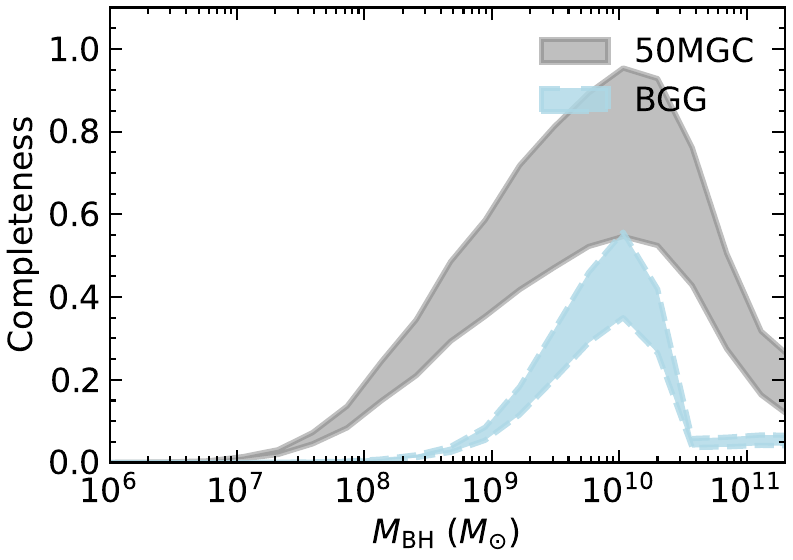}
\caption{Completeness in stellar mass of our parent samples for the X-ray and radio constraints based on the 50 Mpc Galaxy Catalog (50MGC; \citealt{Ohlson2024}), and optical variability based on the NASA-Sloan Atlas catalog of $z<0.055$ spectroscopically confirmed galaxies with stellar masses $M_{\star} \lesssim 2\times10^{10} M_{\odot}$ from SDSS (BGG;  \citealt{Baldassare2020}).
\label{fig:complete}}
\end{figure}

We determine the incompleteness in stellar mass empirically following the procedure outlined in \citet{Ohlson2024}. First, we randomly sample $N$ galaxies from the \cite{Driver2022} GSMF within a volume of $4/3 \pi (50\ {\rm Mpc})^3$. The \cite{Driver2022} GSMF is derived from data release 4 of the Galaxy And Mass Assembly (GAMA) survey and is well-constrained down to $M_{\star} \sim 10^{6.75} M_{\odot}$. We add an additional random scatter of $0.3$ dex to mimic uncertainty in the stellar mass estimates \citep{Conroy2009}. Then, we divide the number of galaxies in the 50MGC catalog (after excluding galaxies in Virgo, Fornax, or within $10^{\circ}$ of the Milky Way plane) by the expected number of sources in bins of stellar mass. The resulting completeness is shown in Figure~\ref{fig:complete}. We will interpolate this curve to determine the incompleteness weighting factor $p(\bm{I}{=}1 | \bm{x})$ in the likelihood.

\subsubsection{Contribution from X-ray binaries}

\begin{figure}[ht!]
\centering
\includegraphics[width=0.45\textwidth]{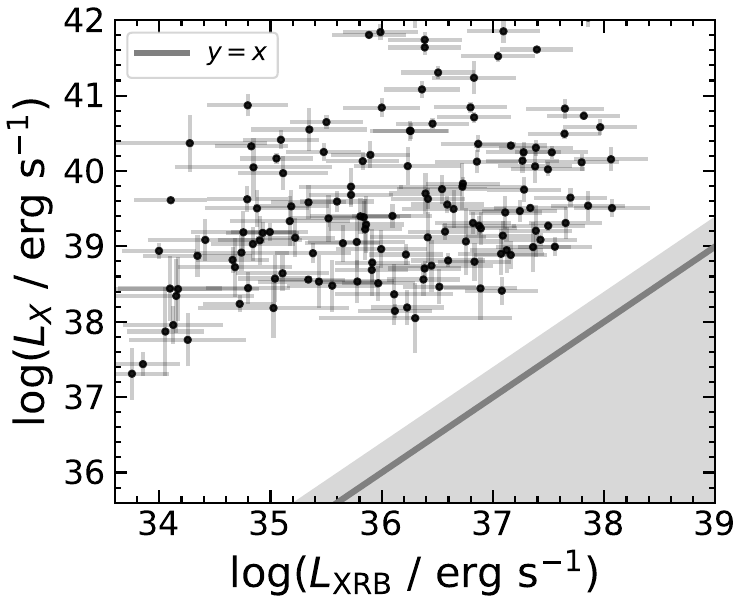}
\caption{Observed nuclear X-ray luminosity versus expected contribution from high and low-mass X-ray binaries within the central aperture in the galaxy corresponding to our matching radius. The expected nuclear X-ray luminosity is estimated by taking the galaxy-wide X-ray luminosity relation from \citet{Lehmer2010} and assuming the X-ray binary population density scales with surface density of the of the galaxy. The gray $y=x$ line is shown plus a shaded X-ray binary region below the +0.4 dex scatter in the \citet{Lehmer2010} relation.
\label{fig:SFR}}
\end{figure}

We model the contribution from XRB populations following \citet{Lehmer2010}: 
\begin{equation} \label{eq:LXRB}
\left( \frac{L_{\rm{XRB}} }{ {\rm{erg}}\ {\rm{s}}^{-1}} \right) = \alpha_{\rm{XRB}} \left( \frac{M_{\star}}{M_{\odot}} \right) + \beta_{\rm{XRB}} \left( \frac{ {\rm{SFR}} }{ M_{\odot}\ \rm{yr^{-1}}}  \right) \\
\end{equation}
where $\alpha_{\rm{XRB}} = (9.05 \pm 0.37) \times 10^{28}$ and $\beta_{\rm{XRB}} = (1.62 \pm 22) \times 10^{39}$. This relation is the total (galaxy-wide) contribution from low and high-mass XRBs with an intrinsic scatter of $\sim 0.4$ dex. The low-mass XRB population is generally more important for early-type galaxies and scales with stellar mass. The high-mass XRB population is generally more important for late-type galaxies and scales instead with the SFR. However, even in late-type galaxies, the bulge is dominated by low-mass XRBs \citep{Mineo2012}. The \citet{Lehmer2010} relation was measured for a sample of galaxies with stellar masses within the range $10^{8} < M_{\ast}/M_{\odot} < 10^{11}$ and star formation rates within the range $-2 < $ SFR/$ M_{\odot}$~yr$^{-1} < 3$. By restricting to unresolved detections, we have ignored the extended emission from hot gas \citep{Mineo2012b}.

We are interested only in the galaxy nuclear X-ray luminosities. As noted by \citet{Miller2015}, a $1^{\prime\prime}$ search radius corresponds to a projected size of about 150 pc at a distance of 30 Mpc. If one were to adopt the typical stellar mass of the nuclear star cluster \citep{Georgiev2016} rather than the stellar mass of entire galaxy, the expected low-mass XRB luminosity would be $\sim 3$ orders of magnitude lower. However, this assumes nuclear star clusters have similar stellar populations as low-mass XRBs in other environments.

We crudely estimate the XRB population within the X-ray search radius by assuming a galaxy-type dependent Sérsic profile and calculating the fraction of galaxy light within in the nucleus of the galaxy (Equation~\ref{eq:nucleation}). We use the \texttt{best\_type} values to determine the galaxy type. We use the \texttt{d25} values for the total galaxy diameter from the 50MGC catalog. This is the 25 mag arcsec$^{-2}$ isophot semi-major axis. When \texttt{d25} values are not available, we assume the size-mass relation based on low-redshift SDSS galaxies from \citet{Shen2003} to estimate the galaxy diameter (in kpc), assuming the total galaxy diameter is 4 times the half-light radius. We then calculate the total angular size from the distance given in the 50MCG catalog. Separate size-mass power law relations are provided for early and late type galaxies by \citet{Shen2003}. We adopt these separate fits depending on the galaxy types given by the \texttt{best\_type} values in the 50MGC catalog. The Sérsic index is also assigned based on galaxy type. These parameters are used to roughly estimate how nucleated the galaxy is in order to estimate the XRB population density within the X-ray search radius following Equation~\ref{eq:nucleation}.

The SFRs are estimated from the infrared (IR) luminosity following \citep{Kennicutt2012,Hao2011,Murphy2011}:
\begin{multline} \label{eq:sfr}
 \log\left(\frac{{\rm{SFR}}}{{M_{\odot}\ \rm{yr}}^{-1}}\right) = \log \left(\frac{L_{\rm{(FUV)}_{\rm{corr}}}}{\rm{erg\ s}^{-1}}\right) - 43.35 \\
 \log L_{{\rm{(FUV)}}_{\rm{corr}}} = L_{{\rm{(FUV)}}_{{\rm{obs}}}} + 3.89\ L_{25\ \mu{\rm{m}}}.
\end{multline}
We use the WISE W4 band (22 $\mu$m) luminosity in place of the IRAS 25 $\mu$m luminosity. The errors introduced by this substitution are much smaller than the $\sim 0.4$ dex scatter in the relationship given by Equation~\ref{eq:LXRB} \citep{Latimer2019}. We caution that the SFRs may be over-estimated from contributions the UV and/or IR emission from the AGN light for the more luminous AGNs. The observed X-ray luminosities are shown in comparison to the expected contribution to the X-ray luminosities from XRBs in Figure~\ref{fig:SFR}. All of our sources have nuclear X-ray luminosities in excess of the expectations from XRB populations. Sources with X-ray luminosities consistent with XRB origin are self-consistently decoupled from the BHOF constraints in our statistical model. For sources lacking both GALEX FUV and WISE 22 $\mu$m detections, we assign a SFR based on the star forming main sequence relationship from 258 low-mass local volume galaxies of \citet{Dale2023}.
We assume uncertainties on these SFRs of $0.3$ dex, reflecting the typical intrinsic scatter in the relations. Non star-forming galaxies fall below this relation. Therefore, we expect these SFRs to be approximate upper-limits, resulting in a conservatively-high estimate for the high-mass XRB population. 

\subsection{Radio data}

We matched the The Karl G. Jansky Very Large Array Sky Survey (VLASS; \citealt{Gordon2021}) epoch 1 quick look catalog to the 50 Mpc Galaxy Catalog (50MGC; \citealt{Ohlson2024}) using a 1.5$^{\prime\prime}$ search radius. VLASS is a 3 GHz radio survey north of $-40^{\circ}$ declination. The VLASS quick look astrometry is typically accurate to $\sim 0.25^{\prime\prime}$. In an attempt to minimize contamination from extended star-formation related emission, we restricted to compact (unresolved) radio sources with de-convolved major axis sizes of $\Psi < 0''.5$. This is comparable to the VLASS minimum de-convolved component size scale of $\Psi \sim 0''.3-0''.4$ and is identical to the compactness definition used by \citet{Gordon2021}. We removed sources with incorrect spectroscopic redshifts as described above. In addition to removing sources within $10^{\circ}$ of the galactic plane and within the Fornax and Virgo clusters, we also restricted the parent sample to sources with declination north of $-40^{\circ}$. VLASS is a 2-4 GHz radio survey of the dec $> -40^{\circ}$ sky. We use the 3 GHz radio luminosities and estimate upper-limits as $3\times$ the typical VLASS RMS sensitivity limit, $\approx 3 \times 2.1\ \mu$m beam$^{-1}$ \citep{Gordon2021}. We convert the radio flux to luminosity using, 
\begin{equation}
L_\nu = 4 \pi d^2 S_{\nu},
\end{equation}
where $d$ is the distance given in the 50MGC catalog. We have ignored the $K$-correction for a given radio spectral index, which is small at these distances of $< 50$ Mpc. We have not attempted to correct for the $\sim 15\%$ level systematically under-estimated flux \citep{Gordon2021}, which will not affect the BHOF inference. More detailed radio measurements will be left for future work.

\begin{deluxetable*}{lcccccccccccc}[t!]
\label{tab:radio}
\tablewidth{0.98\textwidth}
\tablecaption{Galaxies with compact radio detections in the VLASS quick look epoch 1 catalog}
\tablehead{
\colhead{\texttt{objname}} & \colhead{\texttt{ra}} & \colhead{\texttt{dec}} & \colhead{\texttt{type}} & \colhead{\texttt{dist}} & \colhead{\texttt{e\_dist}} & \colhead{\texttt{logmass}} & \colhead{\texttt{e\_logmass}} & \colhead{\texttt{logLR}} & \colhead{\texttt{e\_logLR}} & \colhead{\texttt{logL\_IR}} & \colhead{\texttt{e\_logL\_IR}}  & \colhead{\texttt{p\_{{SF}}}} \\
& \colhead{(deg)} & \colhead{(deg)} & & \colhead{(Mpc)} & \colhead{(Mpc)} & \colhead{($\log M_{\odot}$)} & \colhead{(dex)} & \colhead{($\log$ erg s$^{-1}$)} & \colhead{(dex)} & \colhead{($\log$ erg s$^{-1}$)} & \colhead{(dex)}  & \colhead{} 
}
\startdata
ESO434-040 & 146.9176 & -30.9489 & early & 31.191 & 8.187 & 9.96 & 0.233 & 37.35 & 0.43 & 38.04 & -1.0 & 0.0 \\
ESO443-042 & 195.8739 & -29.8288 & late & 35.696 & 6.347 & 9.967 & 0.156 & 36.68 & 0.43 & 38.04 & -1.0 & 0.0 \\
ESO498-003 & 140.902 & -26.8818 & late & 31.018 & 7.813 & 9.855 & 0.224 & 36.74 & 0.43 & 37.95 & -1.0 & 0.0 \\
ESO506-033 & 190.0575 & -25.3269 & early & 12.117 & 5.816 & 9.693 & 0.454 & 36.24 & 0.43 & 37.8 & -1.0 & 0.0 \\
ESO507-021 & 192.6193 & -26.8426 & early & 40.529 & 7.162 & 10.464 & 0.155 & 37.34 & 0.43 & 38.48 & -1.0 & 0.0 \\
\dots & \dots & \dots & \dots & \dots & \dots & \dots & \dots & \dots & \dots & \dots & \dots
\enddata
\tablecomments{\texttt{objname} - Common galaxy name (\texttt{objname}); \texttt{ra} - Right ascension (\texttt{ra}); \texttt{dec} - Declination (\texttt{dec}); \texttt{type} - Galaxy type (\texttt{best\_type}); \texttt{dist} - Distance (\texttt{bestdist}); \texttt{e\_dist} - Error on the distance (\texttt{bestdist\_error}); \texttt{logmass} - Logarithm of the galaxy stellar mass (\texttt{logmass}); \texttt{e\_logmass} - Error on the logarithm of the galaxy stellar mass (\texttt{logmass\_error}); \texttt{logLR} - Logarithm of the adopted 3 GHz radio luminosity; \texttt{e\_logLR} - Error on the logarithm of the adopted 3 GHz radio luminosity (excluding contribution from distance error); \texttt{logL\_IR} - Logarithm of the IR luminosity from WISE; \texttt{e\_logL\_IR} - Adopted error on the IR luminosity (values of -1 indicate an upper-limit); \texttt{p\_SF} - Posterior probability that the luminosity is due to star-forming process rather than black hole accretion. The names in parenthesis are the corresponding column names from the \citet{Ohlson2024} catalog. These column data are repeated here for completeness. Only a portion of the table is shown here. The full version is available as supplementary data.}
\end{deluxetable*}

\subsubsection{Contribution from star formation processes}

\begin{figure}[ht!]
\centering
\includegraphics[width=0.45\textwidth]{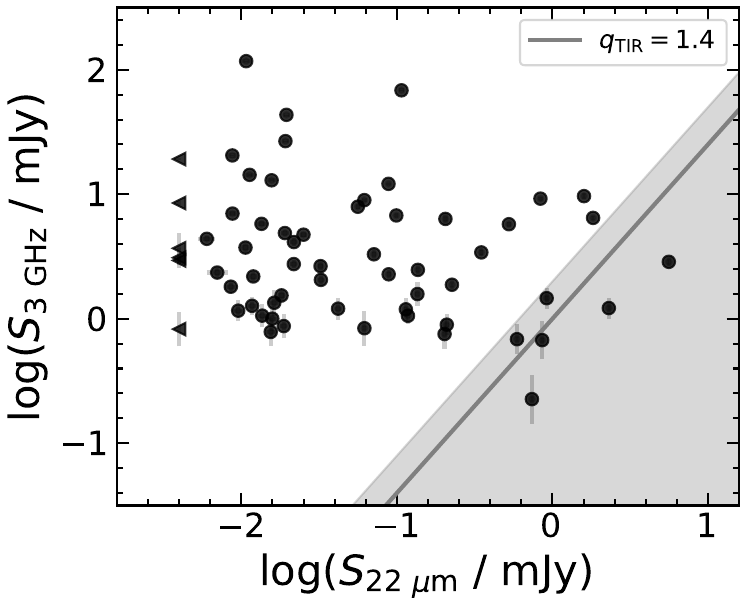}
\caption{Observed compact 3 GHz VLASS radio flux vs. the WISE4 22 $\mu$m infrared flux. Left arrows are upper-limits from the WISE catalog. The expected radio-to-infrared flux ratio $q_{\rm{TIR}}$ from star formation in the galaxy is shown as the $y=x$ line plus a shaded star formation dominated region below the +0.3 dex scatter in the \citet{Appleton2004} relation.
\label{fig:q}}
\end{figure}

We use the infrared-to-3 GHz radio luminosity ratio to distinguish between radio emission from an AGN and star formation (e.g., \citealt{Condon1992,Appleton2004,Delhaize2017,Delvecchio2021}), conservatively assuming that the infrared luminosity is dominated by star formation. This assumption is probably reasonable given the low AGN luminosities we have selected. The infrared-to-1.4 GHz relation from star formation processes from \citet{Appleton2004} is,
\begin{equation}
    \log(S_{24~\mu{\rm m}}/S_{1.4~{\rm GHz}}) \\
    = 0.84,
\end{equation}
with a scatter of $\sim 0.3$ dex. As before, we use the WISE $22~\mu$m flux in place of the $24~\mu{\rm m}$ flux. We convert this ratio to 3 GHz assuming a radio spectral index from star formation of $\alpha = -0.7$ (e.g., \citealt{Condon1992,Appleton2004}):
\begin{equation}
    q_{{\rm TIR}} = \log(S_{22~\mu{\rm m}}/S_{3~{\rm GHz}}) = 1.4.
\end{equation}
The relation is shown in Figure~\ref{fig:q} for our compact VLASS detections.

\subsection{Optical variability data}

\citet{Baldassare2018,Baldassare2020} identified a sample of 135 and 237 AGNs based on optical variability from the SDSS Stripe 82 and Palomar Transient Factory (PTF) surveys, respectively. We use the PTF dataset, which includes a larger number of variable sources at low stellar masses. The parent sample of the \citet{Baldassare2020} PTF variable AGNs is the NASA Sloan Atlas catalog of spectroscopically-confirmed SDSS galaxies. This catalog extends to further distances of $z<0.055$ than the 50MGC ($z\lesssim0.012$). We recomputed the host galaxy stellar masses using the updated spectroscopic redshifts from SDSS data release 16 or GAMA wherever available (Appendix \ref{sec:cigale}). 

We use the $R$ band light curve root mean square (RMS) from \citet{Baldassare2020} for AGN-like variables and non-variable galaxies after removing a small number of sources with incorrect redshifts by cross checking with updated spectroscopic redshifts from SDSS DR16 \citep{Ahumada2020}. The parent sample includes a cut with stellar masses $M_{\star} < 2\times10^{10} M_{\odot}$. We convert the $R$ band RMS variation in magnitudes to luminosity $\Delta L_{R}$ using the aperture flux given by \citet{Baldassare2020}. We treat this quantity as a proxy for AGN luminosity given the known correlation between AGN amplitude and luminosity/mass \citep{Kelly2009,MacLeod2010} of,
\begin{equation}
\log(L_{R} / {\rm erg\ s}^{-1}) \propto - \log {\rm{BC}} - 3.4 \log(\Delta R),
\end{equation}
where $\Delta R$ is the RMS variation in magnitudes, and we have substituted the bolometric correction conversion of $M_i = 90 - 2.5 \log(L_{\rm{bol}}/ {\rm erg\ s}^{-1})$ \citep{Shen2009} into the best-fit relation from \citet{Suberlak2021}. The scatter in this relation will be folded into the intrinsic scatter in the normal distribution that is convolved with the intrinsic ERDF (Equation~\ref{eq:erdf}). We use the RMS of non variable galaxy light curves (i.e, of the photometric noise in the light curves) as upper limits on the optical AGN luminosity $L_{R}$. We caution that the aperture luminosities include a portion of host galaxy light. The fractional contribution of host galaxy to AGN continuum is larger for low-luminosity AGNs \citep{Shen2011}. Therefore, we chose a normalization luminosity that roughly matches the optical AGN LF. We urge caution when interpreting the exact values of the resulting optical ERDF, which depend on these assumptions. However, the exact conversion to AGN luminosity will not substantially alter the occupation fraction constraints as long as the luminosity detections and non-detections are calculated consistently.


\begin{figure*}[htp!]
\centering 
\includegraphics[width=0.98\textwidth]{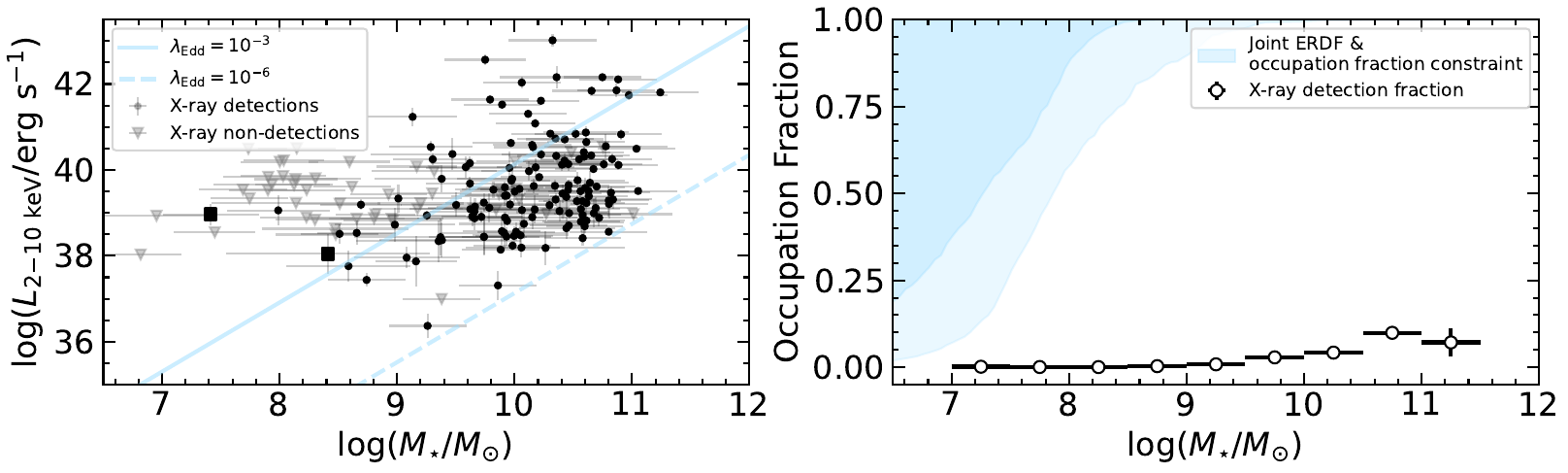}

\includegraphics[width=0.98\textwidth]{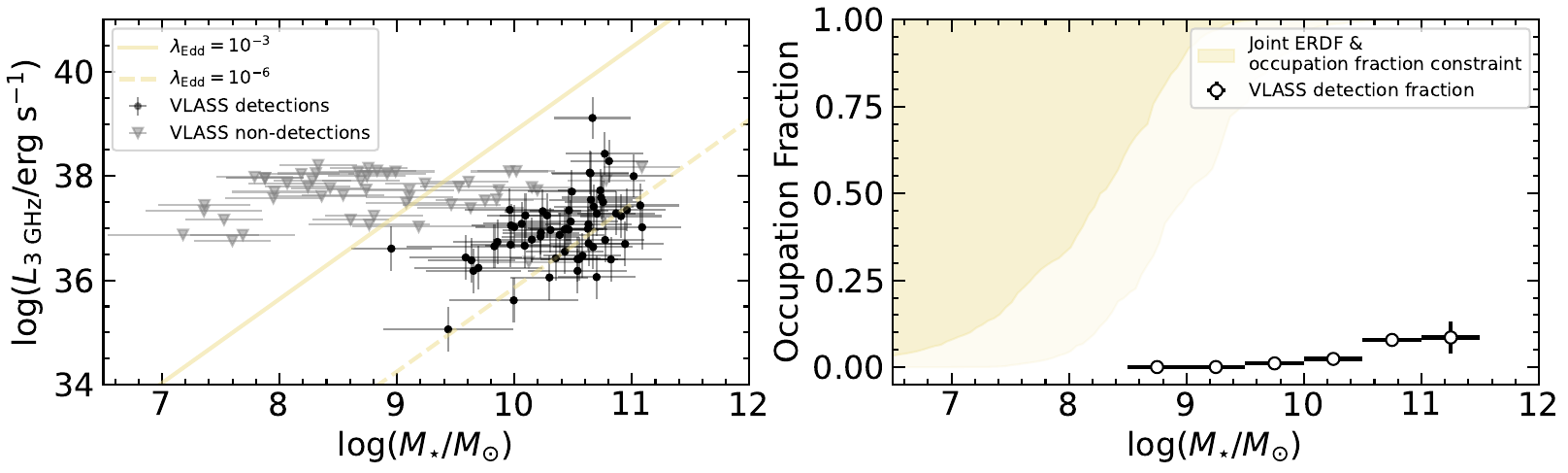}

\includegraphics[width=0.98\textwidth]{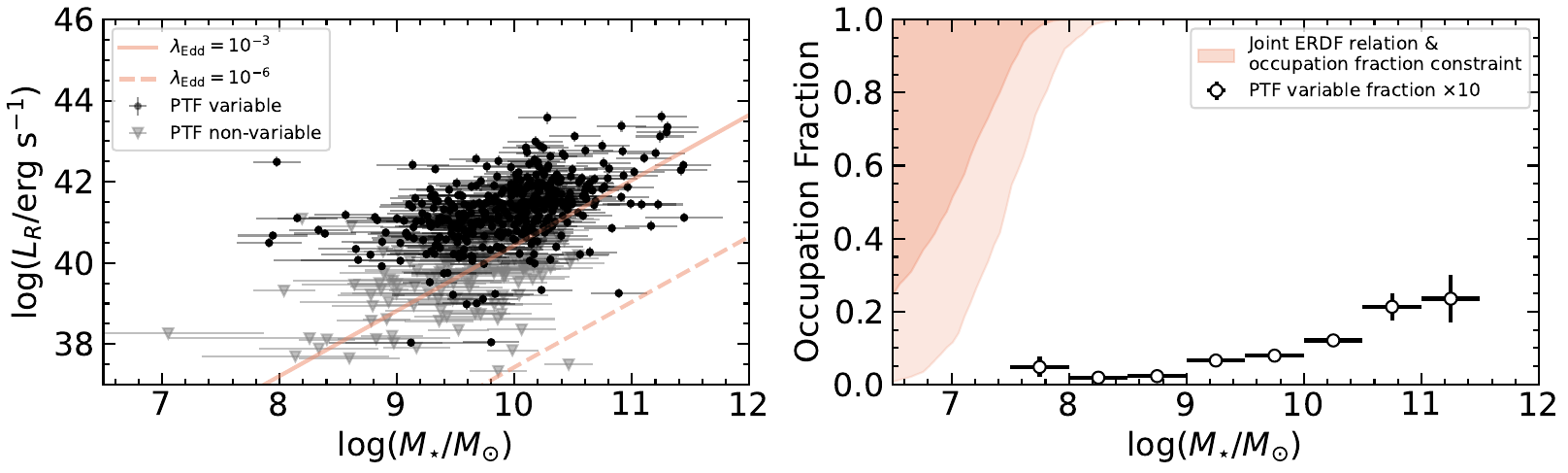}
\caption{Results from individual fitting of X-ray  ($< 50$ Mpc), radio ($< 50$ Mpc), and optical variability  ($z < 0.055$) datasets. \emph{Left panels}: AGN luminosity versus galaxy stellar mass for the X-ray (\emph{upper panel}), radio (\emph{middle panel}), and optical variability (\emph{lower panel}) datasets. Detections are shown as black circle symbols and non-detections are gray triangle symbols along with $1\sigma$ uncertainties on the quantities. A random subset of the non-detections are shown to avoid crowding the figure. The solid and dashed lines are lines of constant Eddington ratio, $\lambda=10^{-3}$ and $\lambda=10^{-6}$, respectively. The data are assumed to be distributed following a smoothly broken power law perpendicular to a line of constant Eddington ratio. The black square symbols are data that have a greater than 5\% posterior probability of being due to an XRB rather than an AGN. \emph{Right panels}: The data-points are the binned detection fractions as a function of stellar mass with 1$\sigma$ uncertainties in the $y$ direction estimated assuming a binomial probability. The darker and lighter shaded regions are the 1$\sigma$ and 95$\%$ credible intervals on the black hole occupation fraction, respectively.
\label{fig:inference}}
\pagebreak
\end{figure*}

\begin{figure*}[ht!]
\centering 
\includegraphics[width=0.98\textwidth]{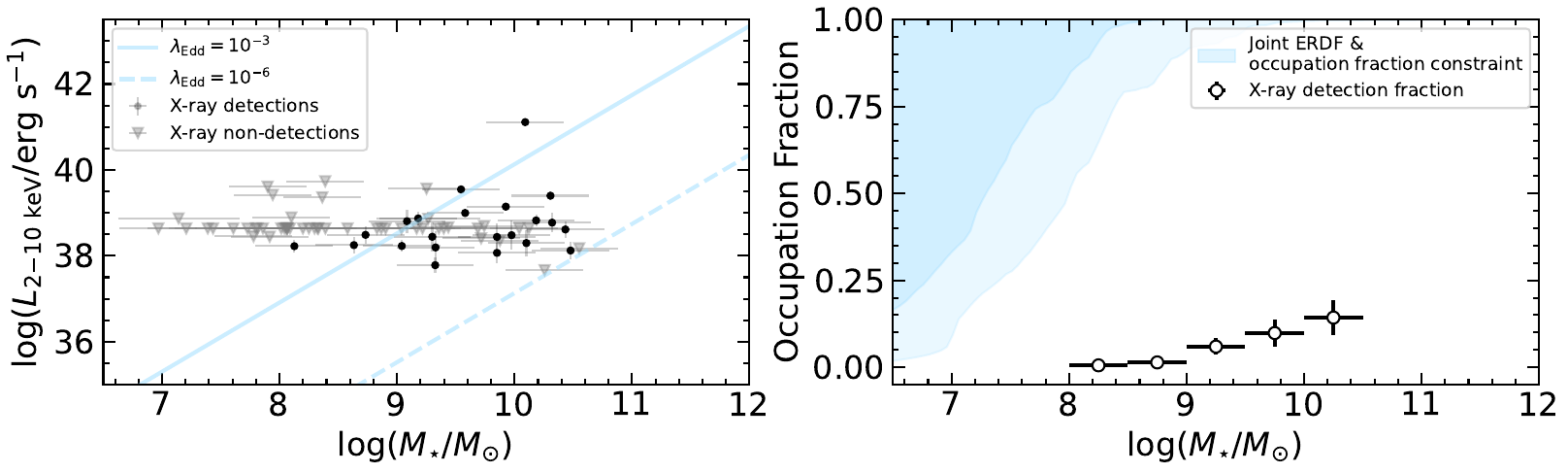}
\caption{Same as Figure~\ref{fig:inference} but for the X-ray data in the Virgo Cluster. Compared our 1$\sigma$ constraints for field galaxies, our results in the Virgo Cluster are consistent within the 1$\sigma$ limits.}
\label{fig:virgoinference}
\end{figure*}

\section{Resulting Occupation fraction constraints} \label{sec:result}

\subsection{Individually-fitting the X-ray, radio, and optical variability dataset}

Our inferred BHOF constraints derived from our collected X-ray, radio, and optical variability data are shown in Figure~\ref{fig:inference}. In every case, our constraints are consistent with full occupation. We are unable to place a more stringent upper limit on the BHOF given the sensitivity of our data. However, we place stringent lower-limits on the BHOF with each dataset. We show our inferred limiting contours on the BHOF that correspond to 1$\sigma$ and 95\% credible intervals. Our constraints fully account for the uncertainty in the shape and scatter in the ERDFs. We will show the resulting ERDFs and luminosity functions in \S\ref{sec:erdf} and \ref{sec:LF}. To facilitate comparison with previous work, we will typically quote the stellar mass where the occupation fraction is at least 50\%.

The optical variability data is the most constraining, requiring full occupation in galaxy stellar masses of $M_{\star} \gtrsim 10^{8} M_{\odot}$. The optical variability data requires an occupation fraction of at least 50$\%$ at $M_{\star} \sim 10^{7.5} M_{\odot}$. Meanwhile, the X-ray and radio data are fully consistent with the constraints from optical variability. The X-ray data requires a BHOF of at least 50$\%$ at $M_{\star} \sim 10^{8} M_{\odot}$ due to the shallow depth of VLASS. The radio data places the weakest limits on the BHOF of at least 50$\%$ at $M_{\star} \sim 10^{7.5} M_{\odot}$. This is not surprising given the shallow depth of VLASS radio data that limits the numbers of radio detections in dwarf galaxies.

We caution that the optical variability data is somewhat dependent on the Type 1 fraction. Our assumed functional form of the Type 1 fraction \citep{Oh2015} breaks the degeneracy between with the active fraction and the ERDF. Our results for all datasets at low stellar masses are somewhat sensitive to the reliability of stellar mass estimates as well. However, the general agreement between the independently-fitted data and the consistency of the resulting luminosity functions (\S~\ref{sec:LF}) is reassuring, and it indicates our model of the star-forming related contamination (in the radio and X-ray case) and Type 1 fraction (in the optical variability case) is likely to be reasonable. Further work is needed beyond the scope of this work to confirm the AGN nature of the individual galaxies and to obtain more reliable black hole mass estimates. We discuss implications for seeding models in \S\ref{sec:discussion}.

Our inferred occupation fraction from optical variability is consistent with previous work using forward modeling that found that the heavy and light seeding scenarios were indistinguishable using a binned detection fraction with the \citet{Baldassare2020} PTF dataset \citep{Burke2023}. However, our inferred limit on the occupation fraction in this work is much more constraining. We attribute this difference to improvements in our methodology, which takes into account the probability of every data point rather than binned average detection fractions.

\subsection{X-ray results in the Virgo Cluster}

Motivated by predictions that the occupation fraction differs substantially local field galaxies \citep{Tremmel2023}, we repeat the analysis using a sub-sample of the X-ray dataset in the Virgo Cluster. We restrict the galaxies within 12$^{\circ}$ of the center of the Virgo Cluster and with distances of $|D - 16.5\  {\rm{ Mpc}}| < 3\sigma_D$, where $D$ and $\sigma_D$ are the distance and distance error respectively in Mpc from the 50MGC catalog. The distance of the Virgo Cluster is $16.5\pm0.1$ Mpc \citep{Mei2007}. The result is shown in Figure~\ref{fig:virgoinference}. The inferred 1$\sigma$ BHOF constraints overlap almost entirely with the results from field galaxies in Figure~\ref{fig:inference}, with no discernible difference between the results from field galaxies and in the Virgo cluster. However, the \citet{Tremmel2023} model does not include a light seeding prescription. If light seeding mechanism is invoked to explain our high occupation fraction results, then it is perhaps not surprising that the enhancement in the occupation fraction from more efficient heavy seeding in dense environments is washed-out.

\subsection{Combined multi-wavelength constraints}

Assuming the X-ray, radio, and optical variability are independent experiments, we can perform a combined, self-consistent multi-wavelength analysis with greater statistical power. The combined log likelihood is given by the sum of the log likelihoods:
\begin{equation}
    \log \mathcal{L}_{\rm{combined}} = \log \mathcal{L}_{\rm{X}} +  \log \mathcal{L}_{\rm{radio}} +  \log \mathcal{L}_{\rm{var}},
\end{equation}
where have chosen to weigh each experiment equally. We assume a single, universal BHOF (parameters $M_{\rm{50}}$, $\delta$, $\theta$) while the remaining parameters governing the ERDFs are fitted separately for each dataset. This assumption is motivated by the fact that the BHOF limits are consistent between all three datasets when analyzed independently (Figure~\ref{fig:fbhof}). After constructing the combined likelihood, the posterior distribution sampling then proceeds as usual. 

Although we used different parent samples for the X-ray/radio and optical variability data, it is valid to combine the likelihoods as long as we account for the differing incompleteness in the parent samples $p(I=1|\log M_{\star})$ and are not looking for trends with redshift. The volumes from the parent sample of the radio and X-ray data ($\sim 10^6$ Mpc$^{3}$) and optical variability data ($\sim 10^7$ Mpc$^{3}$) differ substantially, so our combined analysis is not sensitive to possible redshift evolution after $z \sim 0.055$.  Our resulting posterior distribution for the combined analysis is shown in Appendix~\ref{sec:post}. 

\subsection{Consistency with the Fundamental Plane of Black Hole Activity} \label{sec:fp}

\begin{figure}[ht!]
\plotone{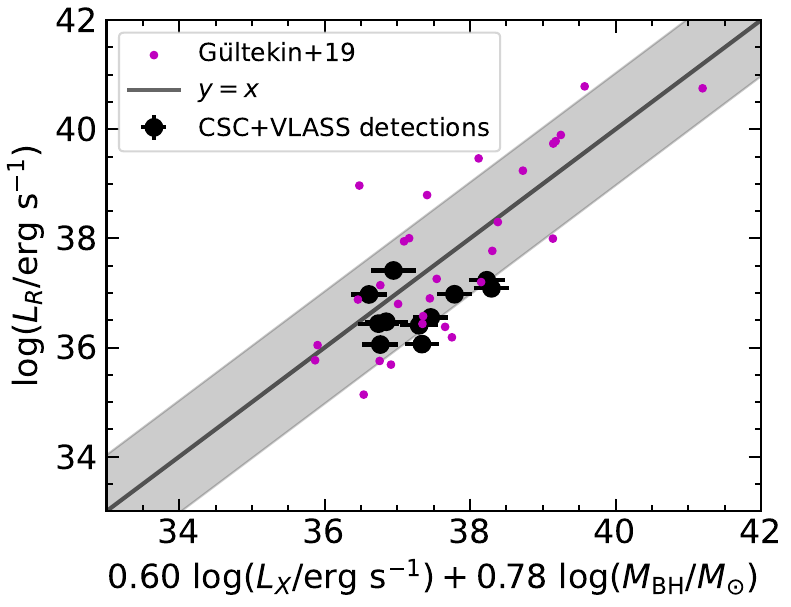}
\caption{Sources with Chandra X-ray and VLASS radio detections on the fundamental plane of black hole activity (black points). The AGNs from the sample of \citet{Gultekin2019} are shown as magenta circle symbols for comparison. The plane is along the $y=x$ line with 1$\sigma$ scatter, given by the quadratic sum of scatter in the fundamental plane with black hole mass and the scatter black holemass -- stellar mass relation. The quantity $M_{\rm{BH}} = 0.0025\ M_{\ast}$ is substituted for the black hole masses \citep{Reines2015}. These sources are largely consistent with the fundamental plane.
\label{fig:fp}}
\end{figure}

The empirical correlation between X-ray luminosity, radio luminosity, and black hole mass is referred to as the ``fundamental plane'' of black hole activity \citep{Merloni2003}. The fundamental plane is usually interpreted as evidence for some form of scale-invariant coupling between the disk and the jet. Despite the large scatter in the relation, the fundamental plane is sometimes used as a black hole mass indicator. But the scatter of $\sim 1$ dex is large in the black hole mass direction, and complications can arise from variability and contamination from other sources when high resolution data is not available \citep{Plotkin2012,Gultekin2022}. \citet{Dong2014} measured a steeper fundamental plane for the low/hard states of XRBs and argued that the \citet{Merloni2003} fundamental plane is more suitable for source with radiatively inefficient accretion.

We show our sources with both X-ray and radio detections on the fundamental plane, using the stellar mass as a proxy for black hole mass in Figure~\ref{fig:fp}. We adopt the fit from \citet{Gultekin2019}. The plane is along the $y=x$ line with 1$\sigma$ scatter given by the quadratic sum of scatter in the fundamental plane with black hole mass and the scatter in the local $M_{\rm{BH}}-M_{\star}$ relation of $s^2 = 0.88^2 + 0.55^2$. The consistency with the fundamental plane relation provides further support that the origin of the X-ray and radio emission in the 50MGC sample is due to accreting SMBHs. Furthermore, we will show that the distribution of luminosities are consistent with the local AGN luminosity functions in \S\ref{sec:LF}.

\subsection{The active fraction and the low-Eddington end of the Eddington ratio distribution function} \label{sec:erdf}

\begin{figure}[ht!]
\centering 
\includegraphics[width=0.48\textwidth]{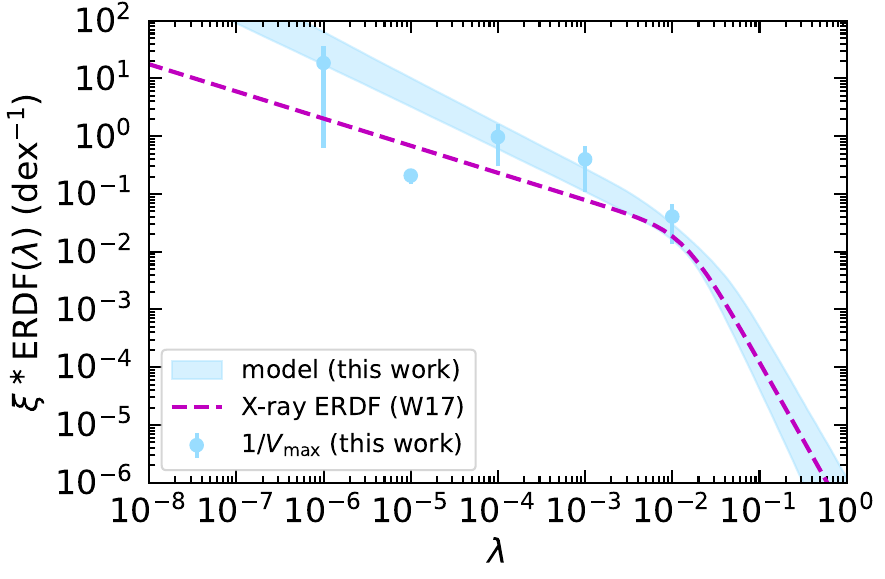}
\caption{The light shaded blue region is our inferred $2{-}10$ keV X-ray ERDF (1$\sigma$ uncertainty band). The blue circle symbols are our binned LF computed with the $1/V_{\rm{max}}$ method. The X-ray ERDFs from previous work are shown for comparison (W17; \citealt{Weigel2017}). }
\label{fig:erdfxray}
\end{figure}

\begin{figure}[ht!]
\centering 
\includegraphics[width=0.48\textwidth]{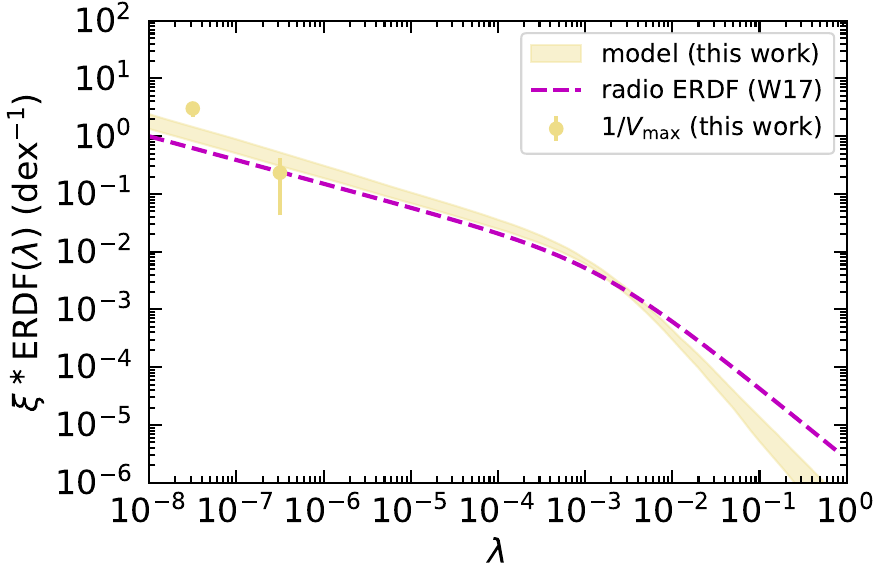}
\caption{The light shaded blue region is our inferred 3 GHz radio ERDF (1$\sigma$ uncertainty band). The yellow circle symbols are our binned LF computed with the $1/V_{\rm{max}}$ method. The radio ERDF from previous work is shown for comparison (W17; \citealt{Weigel2017}).
\label{fig:erdfradio}}
\end{figure}

\begin{figure}[ht!]
\centering 
\includegraphics[width=0.48\textwidth]{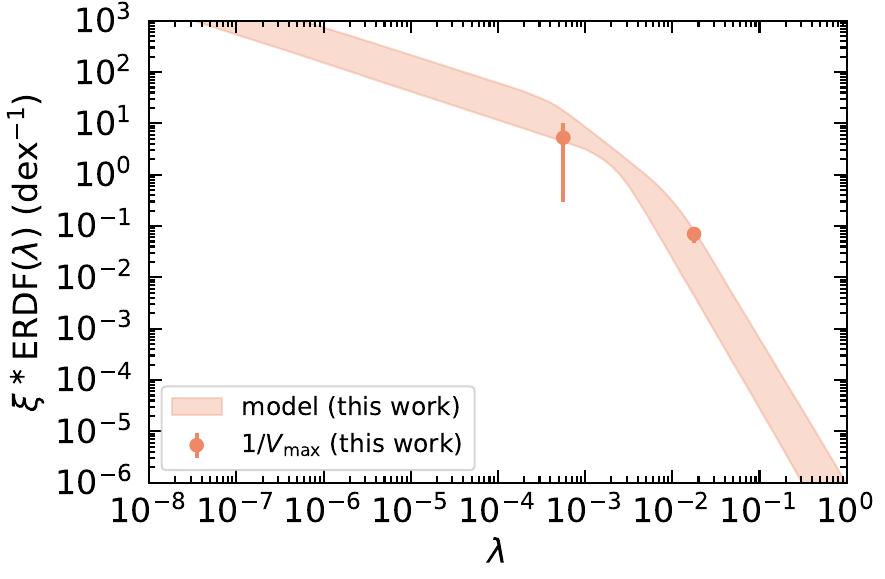}
\caption{The light shaded red region is our inferred $R$ band optical variability ERDF (1$\sigma$ uncertainty band). The red circle symbols are our binned LF computed with the $1/V_{\rm{max}}$ method.
\label{fig:erdfvar}}
\end{figure}

The active fraction refers to the fraction of sources accreting above a chosen Eddington ratio. In our treatment, the active fraction is defined as the fraction of galaxies with Eddington ratios substantially below the minimum Eddington ratio of the ERDF independent of the flux limit. In our fiducial mode, we assume a constant active fraction that is independent of stellar mass. We find an X-ray active fraction of $\sim 15$ percent ($f_a = 0.15 \pm 0.02$), which includes all galaxy types. The radio active fraction is $\sim 50$ percent ($f_a = 0.50 \pm 0.25$). The optical active fraction unconstrained, being somewhat degenerate with the Type 1 fraction. Although our active fractions are consistent between datasets, different active fractions might reflect the intrinsically different ERDFs in the X-ray, radio, and optical sample selections. Furthermore, our definition of active fraction dictates that it is degenerate with the lower bound on the ERDF.

To test whether a mass-dependent active fraction provides a better fit to the data, we modify out model of the active fraction to $f_a \propto A\ (\log M_{\ast} / 8)^c$, (e.g., \citealt{Pacucci2021}). Additionally, we clip the values of $f_a$ to be between 0 and 1. Using this model with the X-ray data for field galaxies, we find no difference (within 1$\sigma$) with the inferred occupation fraction limits compared to a constant, mass-independent active fraction. Our inferred value of $c$ is completely unconstrained within reasonably chosen prior values of $0 < c < 6$ and we find $0.2 < A < 1$, which corresponds to a constant active fraction of at least $\sim 20\%$ independent of stellar mass.

We show our inferred Eddington ratio distribution functions for the X-ray, radio, and optical variability datasets in Figures~\ref{fig:erdfxray}, \ref{fig:erdfradio}, and \ref{fig:erdfvar}. For the X-ray and radio data, we also show the best-fit ERDF from \citet{Weigel2017}. Our X-ray and radio ERDFs are broadly consistent with the mass-independent ERDFs from \citet{Weigel2017}. Although, we lack the detections of rare sources with high Eddington ratios in the radio, the lack of detections still contains some information about the density of the ERDF. Note that the normalization of the ERDFs is degenerate with the integration bounds, i.e., the minimum and maximum Eddington ratio. We will demonstrate consistency with the observed luminosity functions in \S\ref{sec:bhmf}. Taken all together, the model of two mass-independent ERDFs/active fractions for the early and late type galaxy populations appears to be highly consistent with observed data.

\section{Black Hole Mass and Luminosity Functions} \label{sec:bhmf}

\begin{figure*}[ht!]
\centering
\includegraphics[width=\textwidth]{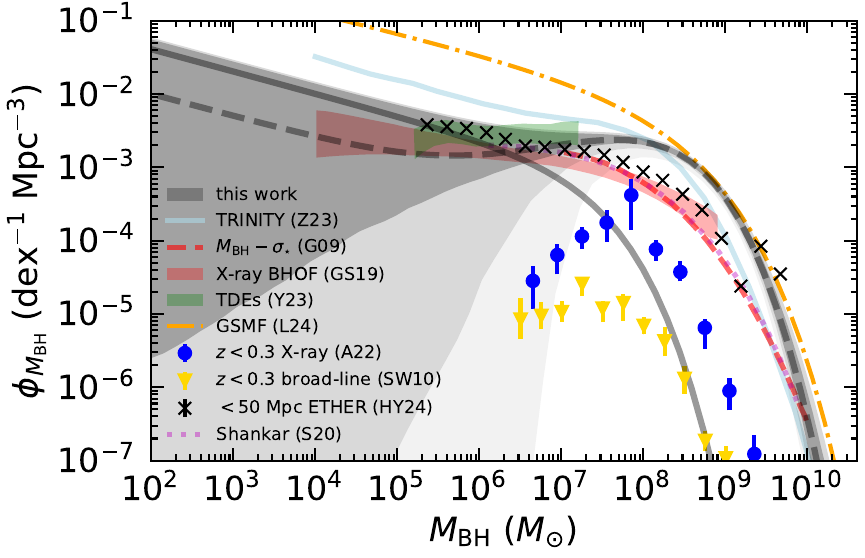}
\caption{Inferred BHMFs from our multiwavelength inference. Our total BHMF for all galaxy types is shown as the gray shaded bands (dark to light: 1, 2, 3$\sigma$ limits). Our late-type BHMF is computed by convolving the product of the BHOF and GSMF for late-type (blue) galaxies with the $M_{\rm{BH}}-M_{\star}$ relation for inactive late type galaxies (solid gray line; \citealt{Greene2020}), which is similar to the AGN relation \citep{Reines2015}. Our early-type BHMF is computed by convolving the product of the BHOF and GSMF for early-type (red) galaxies with the $M_{\rm{BH}}-M_{\star}$ relation for inactive early-type galaxies (dashed gray line; \citealt{Greene2020}). The total (summed) BHMF is shown in the thick dark and light gray bands with $1\sigma$ and 95\% credible interval (CI) bounds, respectively. We show the BHMFs from empirical models (S20; \citealt{Shankar2009,Shankar2020}; Z23; \citealt{Zhang2023}) and from the $M_{\rm{BH}}-\sigma_{\star}$ relation (G09; \citealt{Gultekin2009}), and the ETHER$+$WISE (HY24; \citealt{Hernandez-Yevenes2024}). The observed BHMFs from X-ray AGNs (A22; \citealt{Aird2012}), optical broad-line AGNs (SW10; \citealt{Schulze2010}), TDEs (Y23; \citealt{Yao2023}) with the normalization from (HY24; \citealt{Hernandez-Yevenes2024}) that depends on the TDE rates, and from the GSMF and the \citet{McConnell2013} $M_{\rm{BH}}-M_{\star}$ relation (L24; \citealt{Liepold2024}). 
\label{fig:bhmf}}
\end{figure*}

\subsection{Black Hole Mass Function}

We can derive new constraints on the BHMF assuming the relationship between the black hole mass and stellar mass of the host galaxy and the BHOF. This approach assumes that the active and inactive galaxy populations have the same occupation fractions. Specifically, the BHMF is given by convolving the product of the BHOF and the GSMF with the $M_{\rm{BH}}-M_{\star}$ relation \citep{Caplar2015,Weigel2017,Gallo2019}:
\begin{multline}
\label{eq:BHMF}
    \phi_{\rm{BH}}(M_{\rm{BH}}) = 
    \\ \int^{\log M_{\star, \rm{max}}}_{\log M_{\star, \rm{min}}} (f_{\rm{BHOF}}\ \phi_{M_{\star}})(\log M_{\star})\ \times \\ \mathcal{N}(\log M_{\rm{BH}}-\log M_{\star}|\mu, \sigma_{\star}^2) \ d \log M_{\star},
\end{multline}
where $\mathcal{N}(\mu, \sigma^2)$ a normal distribution with mean $\mu=\log M_{\rm{BH}} - \alpha_{\star} - \beta_{\star} (\log M_{\star} - 10.5)$ and variance $\sigma_{\star}^2$ of the $M_{\rm{BH}}-M_{\star}$ relation depending on galaxy type. We assume an intrinsic scatter of $\sim 0.3$ dex \citep{Liepold2024} for purposes of calculating the BHMFs. We separately calculate the BHMF for early and late type galaxy populations as described below. By multiplying by the BHOF rather than the total detectable fraction $f_d$ in Equation~\ref{eq:BHMF}, our resulting BHMF is corrected for incompleteness and represents the total BHMF, not just the active population.


The local $M_{\rm{BH}}-M_{\star}$ relation for inactive early-type galaxies and inactive late-type galaxies/AGNs are significantly different \citep{Reines2015,Sahu2019,Greene2020}. Early type/elliptical galaxies have systemically larger black hole masses than late-type galaxies/AGNs. This implies different local BHMFs for these two populations. This is not surprising given most AGNs are observed in blue host galaxies with active star formation, while SMBHs with lower accretion rates tend to predominately exist in gas-depleted and quenched red galaxies \citep{Fabian2012,Kormendy2013,Heckman2014,Weigel2016}.

Therefore, we compute separate BHMFs for both early and late type host galaxies as in \citet{Greene2020}. We compute a late-type galaxy BHMF using the late-type $M_{\rm{BH}}-M_{\star}$ relation from \citet{Greene2020} and the red GSMF of \citet{Baldry2012}. Similarly, we compute an early-type galaxy BHMF use the early-type relation from \citet{Greene2020} and the blue GSMF of \citet{Baldry2012}. We also compute a total BHMF using the early-type $M_{\rm{BH}}-M_{\star}$ relation and the total GSMF from \citet{Driver2022}, which is very well constrained to $M_{\star} \sim 10^{6.7} M_{\odot}$. The \citet{Driver2022} GSMF is consistent with the \citet{Baldry2012} GSMF, but \citet{Baldry2012} provide tailored fits to the separate GSMFs using a color cut that is roughly equivalent to separating by early and late types.

Our combined multiwavelength BHMF is shown in Figure~\ref{fig:bhmf}. The inactive early-type BHMF corresponds to the dominant SMBH population ($M_{\rm{BH}} \gtrsim 10^{7} M_{\odot}$). The high mass end of the BHMF is sensitive to the scatter in the $M_{\rm{BH}}-M_{\star}$ relation. The low-mass end of the BHMF is sensitive to the slope of the $M_{\rm{BH}}-M_{\star}$ relation and the shape and uncertainty in the BHOF. For comparison, we show the similarly-derived BHMF of \citet{Gallo2019} which used the \citet{Reines2015} $M_{\rm{BH}}-M_{\star}$ relation for AGNs. We caution that our BHMF is very uncertain below $M_{\rm{BH}} \sim 10^5 M_{\odot}$, where we rely on an extrapolation of the $M_{\rm{BH}}-M_{\star}$ relation. Black holes that are either not in the galaxy centers or with very small masses that fall far below the scatter in the $M_{\rm{BH}}-M_{\star}$ relation will not be counted.

\subsection{Comparison to Observed Black Hole Mass Functions}

The BHMF can be constrained from broad-line $M_{\rm{BH}}$ estimates in AGNs from binning or maximum likelihood methods (e.g., \citealt{Yu2002,Greene2005,Schulze2010,Kelly2009}), from the AGN LF and an Eddington ratio distribution \citep{Hopkins2007,Weigel2016,Ananna2022,Burke2023}, or host galaxy scaling relations (e.g., $M_{\rm{BH}}-\sigma_{\star}$, $M_{\rm{BH}}-L_{\star, {\rm{bulge}}}$, or morphology relations; \citealt{Marconi2004,Lauer2007,Graham2007,Vika2009,Gultekin2009,Davis2014,Mutlu-Pakdil2016}). These observed BHMFs can be compared to results from empirical models \citep{Volonteri2009,Shankar2009,Natarajan2012,Shankar2013,Zou2022}, semi-analytic models \citep{Ricarte2018,Zou2024}, and cosmological numerical simulations \citep{Haidar2022,Tremmel2023}.

We compare our inferred BHMFs using the occupation fraction method to the observed binned BHMFs from the literature. \citet{Schulze2010} measured the $z<0.3$ BHMF from broad-line AGNs from the Hamburg/ESO Survey. Their BHMF is only the Type 1 population, so it falls below the total BHMF from X-ray AGNs \citep{Ananna2022}. \citet{Ananna2022} measured the BHMF for X-ray AGNs from the Swift/BAT AGN Spectroscopic Survey. Both the \citet{Schulze2010} and the \citet{Ananna2022} BHMFs miss the population of inactive SMBHs, mostly on the massive end of the BHMF. The turnover on the low mass end of these BHMFs is due to flux limits. It is difficult to make a direct comparison with these BHMFs, because they do not fold-in the inactive or occupation fraction. However, it is reassuring that they both lie below our total BHMF.

Using 33 tidal disruption events (TDEs) from the Zwicky Transient Facility, \citet{Yao2023} derive a BHMF of $\phi(M_{\rm{BH}}) \propto M_{\rm{BH}}^{-0.25}$ between $M_{\rm{BH}} \approx 10^{5.1-8.2} M_{\odot}$. The normalization of the TDE BHMF is set by the TDE rate, which is very uncertain. We adopt the normalization of \citet{Hernandez-Yevenes2024}. The shape and normalization are highly consistent ($< 1\sigma$) with our BHMF. TDE hosts are strongly biased toward ``green valley'' galaxies \citep{Hammerstein2021} and could therefore perhaps have a $M_{\rm{BH}}-M_{\star}$ in between that of early and late type galaxies.

Our total BHMF is higher than the BHMF from the TRINITY empirical model \citep{Zhang2023} and the empirical model of \cite{Shankar2013} at high masses ($M_{\rm{BH}} \gtrsim 10^{8} M_{\odot}$). At $M_{\rm{BH}} \lesssim 10^{8} M_{\odot}$, the BHMFs from these empirical models tend to over-predict the BHMF derived from the GSMF and the early-type $M_{\rm{BH}}-\sigma_{\star}$ relation. This might not be surprising given TRINITY is only calibrated from the quasar BHMF \citep{Kelly2013}. \citet{2021Pesce} showed that different BHMFs computed from empirical models are systematically larger at low redshifts than observational scaling relations predict.

Our BHMF is consistent with the BHMF from \citet{Liepold2024} above $M_{\rm{BH}} \sim 10^{8} M_{\odot}$. The BHMF from this work and from empirical models are generally higher than the BHMFs derived from other, single scaling relation arguments (e.g., \citealt{Mutlu-Pakdil2016,Hernandez-Yevenes2024}) and the $M_{\rm{BH}}-\sigma_{\star}$ relation \citep{Gultekin2009}. The discrepancy between the predicted high mass end of the BHMF from the $M_{\rm{BH}}-\sigma_{\star}$ relation and the $M_{\rm{BH}}-M_{\star}$ is puzzling and perhaps indicates a selection bias in one or both of the relations. Including the contribution to the BHMF from all galaxy types with the appropriate scaling relations is important across the full black hole mass range.

\begin{figure}[ht!]
\centering
\includegraphics[width=0.48\textwidth]{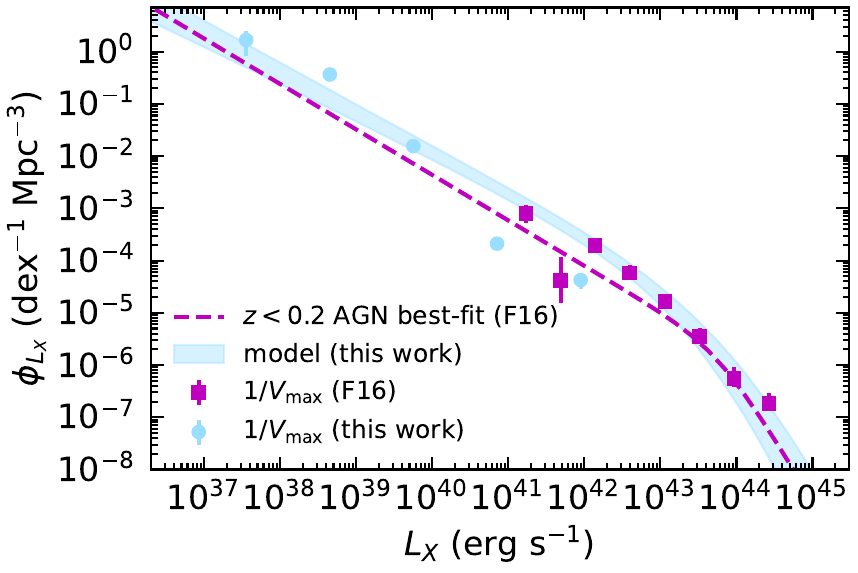}
\caption{The light shaded blue region is our inferred $2{-}10$ keV X-ray LF (1$\sigma$ uncertainty band). Our inferred X-ray LF is computed by convolving our BHMF with the X-ray ERDF. Our inferred X-ray LF corrects for the expected contribution from XRBs assuming the scaling with galaxy SFR \citep{Lehmer2010}. The blue circle symbols are our binned LF computed with the $1/V_{\rm{max}}$ method. The observed X-ray LF from $z<0.2$ AGNs is shown as magenta square symbols (F16; \citealt{Fotopoulou2016}).
\label{fig:xlf}}
\end{figure}

\begin{figure}[ht!]
\includegraphics[width=0.48\textwidth]{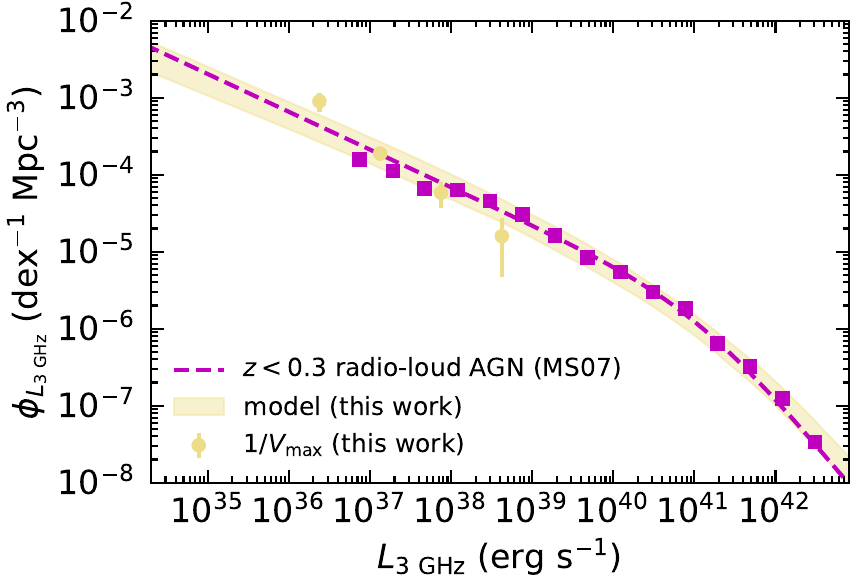}
\caption{The light shaded yellow region is our inferred 3 GHz radio LF (1$\sigma$ uncertainty band). Our inferred radio LF is computed by convolving the product of our BHMF with the radio ERDF. The yellow circle symbols are our binned LF computed with the $1/V_{\rm{max}}$ method. The observed radio LF from $z<0.3$ radio-loud optically-classified AGNs and star-forming galaxies is shown as magenta square and triangle symbols respectively symbols (MS07; \citealt{Mauch2007}).
\label{fig:rlf}}
\end{figure}

\begin{figure}[ht!]
\centering
\includegraphics[width=0.48\textwidth]{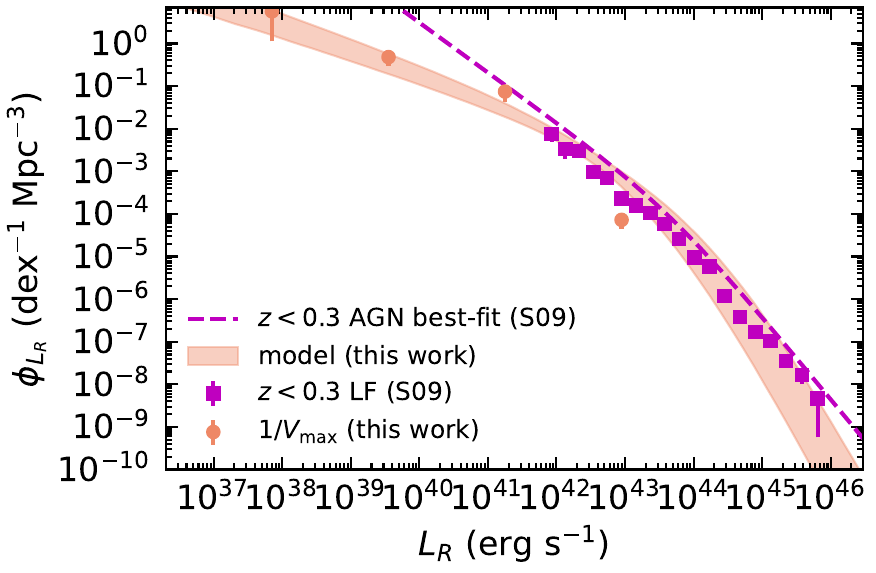}
\caption{The light shaded red region is our inferred $R$ band optical LF (1$\sigma$ uncertainty band). Our inferred LF from optical variability is computed by convolving our BHMF with the optical ERDF and the Type 1 AGN fraction. The red circle symbols are our binned LF computed with the $1/V_{\rm{max}}$ method. The observed optical AGN LF from $z<0.3$ AGNs is shown as magenta square symbols (S09; \citealt{Schulze2009}).
\label{fig:olf}}
\end{figure}

\subsection{Luminosity Functions}
\label{sec:LF}

The AGN LF is given by convolving the product of the BHMF with the ERDF density:
\begin{multline}
\label{eq:LF}
    \phi_L(L) = 
    \\ \int^{\log \lambda_{\rm{max}}}_{\log \lambda_{\rm{min}}} \phi_{M_{\rm{BH}}}(\log L - 38.2 - \log {\rm{BC}} - \log \lambda)\ \times \\ \xi^\ast {\rm{ERDF}}(\lambda) \ d \log \lambda,
\end{multline}
where and $\phi_{M_{\rm BH}}$ is the BHMF, $\xi^\ast$ is a normalization constant that converts the ERDF probability density into a space density, $L$ is the AGN luminosity in a given band. Because our ERDF is a probability distribution function rather than an space density, we simply fit the normalization constant $\xi^\ast$ to the observed LFs using the equation,
\begin{equation*}
    \xi^\ast = \frac{ \phi_{L, {\rm obs}}(L_\ast) }{ \phi_{L}(L_\ast) },
\end{equation*}
where $\phi_{L, {\rm obs}}$ is the observed AGN LF from the literature. This is a required step, because our likelihood function is only measuring the occupation \emph{fraction}. It therefore has no information on the normalization of the mass or luminosity functions.

We are primarily interested in the shape of our LFs and whether they are consistent with the observed AGN LF at $z=0$. We compute LFs for our keV X-ray and 5 GHz radio data. The X-ray LF is shown in Figure~\ref{fig:xlf}. The radio LF is shown in Figure~\ref{fig:rlf}. The optical LF is shown in Figure~\ref{fig:olf}. We checked our inferred LFs using by also computing the LFs with the $1/V_{\rm{max}}$ binning method. Our LFs may be converted into an AGN bolometric LF by assuming an appropriate bolometric correction for low-luminosity AGNs \citep{Duras2020,Lopez2024}. Our adopted observed LFs from the literature are described below.

\subsection{$1/V_{\rm{max}}$ method}

In addition to the parameterized method described above, we also compute the LFs using the $1/V_{\rm{max}}$ method \citep{Schmidt1968} as a consistency check. This approach involves weighing each source by the maximum volume in which it could be detected given its stellar mass. This corrects for Malmquist bias in the sample. The BHMF with $N_{\rm{bin}}$ sources in stellar bin $j$ is computed as:
\begin{equation}
    \phi_{j}\ \Delta \log L = \sum^{N_{\rm{bin}}}_i \frac{w_i}{V_{{\rm{max}}, i}},
\end{equation}
where $1/V_{\rm{max}}$ is the maximum volume in which a source $i$ with a given stellar mass and redshift could be detected and $w$ is a weight corresponding to the reciprocal of the targeting completeness \citep{Weigel2016}. The $1\sigma$ uncertainties on the LF or BHMF $\sigma_{\phi_{j}}$ are estimated as:
\begin{equation}
    \sigma_{\phi_{j}} \approx \sqrt{\sum^{N_{\rm{bin}}}_i \frac{w_i^2}{V_{{\rm{max}}, i}^2}},
\end{equation}
which is valid for large $N_{\rm{bin}}$ \citep{Weigel2016}. We take $w_i^{-1} = p(I_i{=}1|x_i)$ from our previously determined completeness values. We estimated $V_{\rm{max}}$ using the median sensitivity in the CSC or VLASS catalog. We caution that the $1/V_{\rm{max}}$ method is sensitive to binning choices and the minimum stellar mass completeness curve adopted \citep{Weigel2016}.

\subsection{Comparison to Observed Luminosity Functions}

We compare our X-ray LF to the 5$-$10 keV LF from \citet{Fotopoulou2016}. The \citet{Fotopoulou2016} LF was measured from a combination of Chandra and XMM-Newton observations in six deep fields. We use their $0.01 < z <0.2$ LF, which extends to $L_X \sim 10^{41}$ erg s$^{-1}$. We converted the 5$-$10 keV LF from \citet{Fotopoulou2016} to a 2$-$10 keV LF using Equation~\ref{eq:LX}. As before, we assume an X-ray spectral index of $\Gamma=1.8$. The X-ray LF comparison is shown in Figure~\ref{fig:xlf}. Our model LF is highly consistent ($\sim 1\sigma$) with the data points from \citet{Fotopoulou2016} and broadly consistent with an extrapolation of the best-fit \citet{Fotopoulou2016} LF to $\sim 4$ orders of magnitude lower luminosities. The consistency with the AGN LF suggests we are not strongly affected by contamination from XRBs. The low luminosity end of our X-ray LF slope of $dN/d L \propto -0.74\pm0.06$ is also reasonably consistent with the eROSITA X-ray AGN LF slope of $-0.63\pm0.05$ at $<200$ Mpc from \citet{Sacchi2024}.

We compare our radio LF to the 1.4 GHz LF from \citet{Mauch2007}, which is broadly consistent with more recent determinations of the radio LF \citep{Pracy2016} and older work \citep{Filho2006}. The \citet{Mauch2007} LF was measured from Faint Images of the Radio Sky at Twenty-cm survey (FIRST) data. We use their $0.003 < z < 0.3$ LF, which extends to $L_{1.4~{\rm{GHz}}} \sim 10^{36}$ erg s$^{-1}$.  We converted the 1.4 GHz LF from \citet{Mauch2007} to 3 GHz using,
\begin{equation}
L_{3\ {\rm GHz}} = \left(\frac{3}{1.4}\right)^{\alpha} L_{1.4\ {\rm GHz}} .
\end{equation}
with a spectral index of $\alpha=-0.5$ (e.g., \citealt{Zajacek2019}). The radio LF comparison is shown in Figure~\ref{fig:rlf}. Our model LF is highly consistent ($\sim 1\sigma$) with the data points from \citet{Mauch2007}. The consistency with the radio-loud AGN LF suggests we are not strongly affected by contamination from star formation.

We compare our optical LF to the optical AGN LF from \citet{Schulze2009}. The \citet{Schulze2009} LF was measured from the Hamburg/ESO survey at $z <0.3$, which extends to $L_R \sim 10^{42}$ erg s$^{-1}$. We converted their bolometric LF to the optical $R$ band with the same bolometric corrections as before. The optical LF comparison is shown in Figure~\ref{fig:olf}. Our model LF is highly consistent ($\sim 1\sigma$) with the data points from \citet{Schulze2009}, but our model indicates a shallower low luminosity slope than the extrapolated best-fit broken power law of \citet{Schulze2009}. The consistency with the AGN LF suggests we are not strongly affected by contamination from false positive detections at low luminosities.

\begin{figure*}[ht!]
\plotone{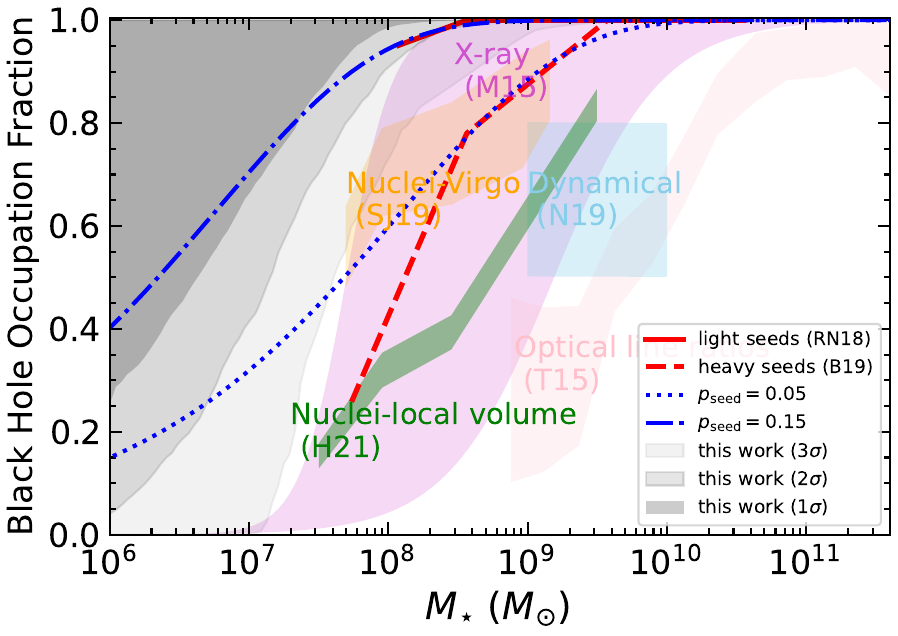}
\caption{Observational constraints on the nuclear black hole occupation fraction as function of host galaxy total stellar mass at $z\sim0$ from our multiwavelength accretion probes (gray shaded regions), using the nuclear star cluster occupation fraction as a proxy for the black hole occupation fraction in the Virgo cluster (SJ19; \citealt{SJ2019}) and local volume galaxies (H21; \citealt{Hoyer2021}), dynamical black hole masses (N19; \citealt{Nguyen2019}), optical narrow emission line ratio diagnostics of $0.01 < z < 0.1$ galaxies (T15; \citealt{Trump2015}), and the AMUSE X-ray data (1$\sigma$ lower limit; M15; \citealt{Miller2015}). Model predictions for heavy and light seeds \citep{Ricarte2018,Bellovary2019} at $z=0$ are overplotted as red solid and dashed lines, respectively. Predictions for varying probabilities that a halo with mass greater than $10^7 M_\odot$ was seeded at high redshift are shown in blue, following an analytic argument based on the statistics of hierarchical growth of halos are shown in blue \citep{Buchner2019} as described in the text.
\label{fig:occupationcombined}}
\end{figure*}

\section{Discussion} \label{sec:discussion}


\subsection{AGNs all the way down?}

We have assumed a simple model that describes the distribution of AGN luminosities at fixed stellar mass as a mass-independent smoothly broken power law, modulated by the black hole occupation fraction. Following \citet{Caplar2015,Weigel2017,Greene2020}, we assume the early and late type galaxies have their respective $M_{\rm{BH}}-M_{\star}$ relations. We have again shown that this mass-independent ERDF corresponding to radio and quasar-mode accretion is highly consistent with observations of the separate radio and X-ray AGN LFs that correct for selection biases. The low-luminosity ends of our inferred LFs are generally consistent with reasonable extrapolations from the AGN LFs derived at higher luminosities. Along with these LFs, the inferred lower limit on the BHOF indicates that every galaxy greater than $M_{\star} \sim 10^{8} M_{\odot}$, and possibly much lower, is likely to contain a central black hole.

The sharp turnover in the ERDFs indicates that while high Eddington ratio AGNs are rare, their prevalence does not decrease with stellar mass in early type galaxies. However, at low masses, we must be lucky to observe an AGN with a high enough Eddington ratio luminosity to overcome the detection limit. Hence, the detectable AGN fraction will always decrease with stellar mass under for a luminosity (or flux) limited sample. The detection rates depend on the shape of the ERDF and scatter in the $M_{\rm{BH}}-M_{\star}$ relation.

On the other hand, \citet{Arcodia2024} note the low eROSTIA X-ray detection rates (17/200) of low-mass AGNs selected from UV/optical/infrared variability. Their stacked X-ray luminosity is $\sim 10^{39}$ erg s$^{-1}$ at $M_{\star} \sim 10^{9.5} M_{\odot}$. Our X-ray luminosities of $\sim 10^{38-41}$ erg s$^{-1}$ are consistent with that of \citet{Arcodia2024} and similar to nearby low-luminosity AGNs \citep{Terashima2003,Ho2009,Williams2022}. As noted by \citet{Arcodia2024}, the X-ray non-detections could also reflect differences in the SED of low-mass AGNs or changes in the X-ray corona. An analysis of the data using a technique similar to our method would be required to determine whether the X-ray ERDF of this sample is consistent with this work. But other sources of variably from e.g., luminous blue variable stars or long-duration supernovae could be contaminating the variable AGN sample and are difficult to rule-out \citep{Baldassare2016,Burke2020lbv} or, in some cases, spurious detections.

\subsection{Comparison to other probes of the occupation fraction} 

We summarize the observational constraints on the occupation fraction at $z\sim0$ in Figure~\ref{fig:occupationcombined}, following \citet{Greene2020,Haidar2022}. We include probes from optical variability (\citealt{Baldassare2020} and this work), X-ray luminosities \citep{Miller2015}, using the nuclear star cluster (NSC) occupation fraction as a proxy for the black hole occupation fraction in the Virgo cluster \citep{SJ2019} and local volume galaxies \citep{Hoyer2021}, X-ray detections in NSCs \citep{Baldassare2022,Hoyer2024}, dynamical black hole masses \citep{Nguyen2019}, and optical narrow emission line ratio diagnostics \citep{Trump2015}. We take model predictions for heavy \citep{Bellovary2019} and light seeds \citep{Ricarte2018} and over-plot these predictions in Figure~\ref{fig:occupationcombined}. \citet{Miller2015} imposed a prior limit of $M_{50} > 10^{7.5} M_{\odot}$. Therefore, we consider their BHOF to be a lower limit consistent with our work (see also Appendix~\ref{sec:amuse}). \citet{Cho2024} inferred a BHOF from broad H$\alpha$ detections consistent with the \citet{Miller2015} results. Their model assumes a constant bolometric correction from AGN bolometric luminosity to broad-line luminosity. However, if the efficiency of photoionizing the broad line region drops at low luminosities or accretion rates \citep{Trump2011,Chakravorty2014,Hagen2024}, then their result could be a lower limit on the true occupation fraction.

We caution that all these observational probes are subject to various selection biases. For example, variability and X-ray and radio data select different populations of galaxies with different accretion modes. NSCs constraints could miss a population of IMBHs that do not coincide with a NSC or overcount NSCs lacking an IMBH (see \citealt{Neumayer2020}), and emission line ratio diagnostics are known to be very inefficient at classifying AGNs in metal-poor and star forming dwarf galaxies \citep{Groves2006,Trump2015,Polimera2022}. We performed no a priori cuts on the host galaxy properties (e.g., type, star formation rate).

All of these methods would miss populations of wandering black holes that are not within the nucleus of the galaxy. Importantly, the dynamical friction timescale suggests that light seeds do not sink to centers of galaxies \citep{Ma2021}. These wandering black holes would have low accretion rates that would make them difficult to detect, though some could manifest as ultra-luminous X-ray sources \citep{DiMatteo2023} or be detectable via TDE flares \citep{Chowdhury2024}. Furthermore, the centers of irregular low-mass galaxies are not well defined and simple catalog cross-matching could result in a lower inferred BHOF at low stellar mass \citep{Ohlson2024}. Although this effect is evidently not substantial given our inferred BHOF lower limit is still consistent with full occupation in all galaxies regardless of stellar mass.

Our occupation fraction constraints are in excess of the occupation fraction derived from the dynamical black hole masses and optical narrow emission line ratios in local galaxies. However, the dynamical detection rates may be dependent on the degeneracies in the Jeans modeling \citep{Greene2020}. It is perhaps worth revisiting the assumptions in the \citet{Trump2015} model with more recent datasets (e.g., \citealt{Polimera2022}). Or perhaps we have vastly under-estimated the contamination from non-AGNs or false positive detections from variability \citep{Arcodia2024}. Our constraints are more consistent with the light seed occupation fraction prediction from \citet{Ricarte2018} than the heavy seeding only predictions \citep{Bellovary2019}. However, the uncertainties on the semi-analytic model predictions are difficult to quantify. Additionally, continuous black hole formation channels in NSCs \citep{Natarajan2021} could result in a BHOF fraction signature at $z\sim0$ that elevates the total BHOF and traces the NSC occupation fraction.

\subsection{Implication for seeding}

Recent observations of very high redshift quasars suggest they likely formed from heavy seeds in the early Universe \citep{Natarajan2024,Andika2024}. Assuming high-redshift SMBHs grew predominately by accretion, \citet{Fragione2023} inferred the initial mass function of black hole seeds. Our comparison with the semi-analytic model results of \citet{Ricarte2018} suggest that their heavy seeding only model is ruled-out by our observations with $>95\%$ confidence. This semi-analytic model and results from numerical simulations \citep{Ni2021,Haidar2022} suggests that is difficult to explain an occupation fraction of $\sim 1$ at $M_{\star} \lesssim 10^{8} M_{\odot}$ without light seeds. However, we stress that these results do not necessarily rule-out the existence of heavy seeding channels as well. Most likely both seeding channels operate in the universe at varying efficiencies. 

\citet{Buchner2019} derived an analytic prediction for the functional form of the $z=0$ BHOF based on the hierarchical growth of halos over cosmic time. We summarize their argument here. The number of progenitor halos greater than some critical halo mass $M_c$ that built a $z=0$ halo of mass $M_{h,z=0}$ is given by \citet{Menou2001}:
\begin{equation}
    N_{h,z=0} = \frac{1}{2} (M_{h,z=0} / M_c)^{3/4}.
\end{equation}
Assuming each halo with mass greater than $M_c$ had an equal chance $p_{\rm{seed}}$ of containing a seed black hole, the $z=0$ BHOF is,
\begin{equation}
    f_{\rm{BHOF}} = 1 - (1 - p_{\rm{seed}})^{\alpha_h N_{h,z=0}}.
\end{equation}
where the factor $\alpha_h < 1$ represents some efficiency of the seed black holes to sink to the center of the galaxy before $z=0$ where they can be detected. We take $\alpha_h = 0.1$ for consistency with \citet{Ricarte2018}.

We convert the $z=0$ halo mass to stellar mass using the stellar -- halo mass relation of \citet{Girelli2020}. We find that for a fiducial $M_c = 10^7 M_{\odot}$, our results from Figure~\ref{fig:occupationcombined} imply a seed probability of $p_{\rm{seed}} > 0.15$ (95\% CI). We confirm that re-performing the combined multi-wavelength inference with this BHOF yields the same result. Therefore, according to this model, at least 15\% of all halos with mass greater than $10^7 M_{\odot}$ must have been seeded at high redshift. On the other hand, \citet{Lodato2006} expect that $\sim 5\%$ of halos with mass $\sim 10^{7} M_{\odot}$ would be seeded by direct collapse. Selecting $p_{\rm{seed}}=0.05$ and $M_c = 10^{7} M_{\odot}$ is a reasonable match to the semi-analytic model predictions for heavy seeds from \citet{Ricarte2018}, as shown in Figure~\ref{fig:occupationcombined}. However, the value of $p_{\rm{seed}}$ can be much lower if we lower the critical mass threshold $M_c$ that count toward the total number of progenitor halos or increase $\alpha_h$. This is because only a small fraction of progenitor halo needs to have a seed black hole on average for its final $z=0$ halo to be occupied. As a final note, the true seeding probability could me much higher if only a fraction of seeds grow to become massive enough to fall near the $M_{\rm{BH}} - M_{\ast}$ relation.

\subsection{Implications of our local black hole number density constraints}

Our self-consistently derived BHMF is larger in number density at the low mass end than the predictions from AGNs and some empirical models (Figure~\ref{fig:bhmf}). At the high mass end, our results agree will with \citet{Liepold2024}. Our high SMBH space densities are more consistent with the characteristic amplitude of gravitational waves from pulsar timing arrays, as demonstrated by \citet{Liepold2024}. Our elevated mass function normalization at the low mass end of Figure~\ref{fig:bhmf} indicates an enhanced expected amplitude for LISA in the IMBH regime as well. We again caution that our BHMF is a very uncertain extrapolation below $M_{\rm{BH}} \sim 10^{4} M_{\odot}$. More work is needed in order to understand the origin of the discrepancies in the BHMFs and the selection biases at play. 

It is important to establish alternative ways to constrain the BHMF as a prerequisite to creating reasonable models for gravitational wave observations such as pulsar timing arrays \citep{Agazie2023} and LISA \citep{Amaro-Seoane2017}. Although uncertainties in SMBH merger rates are very large and their mergers are expected to be more common at higher redshifts, a consistent BHMF from observations and empirical and semi-analytic models in the local universe would have implications for their local merger rate expectations. In addition, comparing the BHMFs derived from host galaxy scaling relations to other observed BHMFs can constrain possible redshift evolution in the scaling relations \citep{Matt2023}.

\citet{Buchner2019} investigated LISA merger rates based on the previous occupation fraction limits from \citet{Miller2015}. The $z=0$ BHMF is the integral of mass assembled over cosmic time from accretion and mergers. Central black holes are assumed to trace halo mergers with some delay function (save those black holes that do not sink to the centers of the galaxies/halos). \citet{Buchner2019} show that the number of LISA events traces the black hole number density, and reflects their initial seeding conditions. Differences in black hole merger delay function systematically shifts the redshift distribution of the LISA events. Using our BHMF from Figure~\ref{fig:bhmf}, we can estimate the $z=0$ black hole number density at $M_{\rm{BH}} > 10^4 M_{\odot}$ in Mpc$^{-3}$ following e.g., \citet{Graham2007}, as,
\begin{equation}
    n_{\rm{BH}} = \int_{10^4 M_{\odot}}^{10^9 M_{\odot}} \phi_{M_{\rm{BH}}}\ d\log M_{\rm{BH}},
\end{equation}
and mass density in $M_{\odot}$ Mpc$^{-3}$,
\begin{equation}
    \rho_{\rm{BH}} = \int_{10^4 M_{\odot}}^{10^9 M_{\odot}} \phi_{M_{\rm{BH}}}\ M_{\rm{BH}}\ d\log M_{\rm{BH}}.
\end{equation}
We find $n_{\rm{BH}} > 10^{-2.6}$ Mpc$^{-3}$ and $\rho_{\rm{BH}} > 10^{4.6}$ $M_{\odot}$ Mpc$^{-3}$ using our 95\% credible interval limits on the BHMF. Assuming LISA is sensitive to all these events, we expect at least a few LISA events per year from coalescing central black holes with $M_{\rm{}} > 10^4 M_{\odot}$ according to the \citet{Buchner2019} model.



\subsection{What about other AGN selection methods?}

In principle, any tracer of AGN activity that scales with the stellar mass of the host galaxy could be used to infer the BHOF using the formalism developed here. The broad emission line luminosity scales with the AGN bolometric luminosity \citep{Greene2005} with a scatter at fixed black hole mass determined by the scatter in the ERDF. \citet{Reines2013,Chilingarian2018,Liu2019,Salehirad2022} identified dwarf galaxies showing broad-line features consistent with AGN activity. Unfortunately, the optical broad-line sources without narrow-line signatures of AGNs are strongly contaminated by supernova emission and stellar winds or transients with broad Balmer lines at low luminosities with highly uncertain rates \citep{Izotov2007,Baldassare2016,Burke2020}. \citet{Baldassare2016} found that 14/16 of the \citet{Reines2013} broad-line AGNs had ambiguous or transient broad emission, indicating those sources are unlikely to be AGNs. Furthermore, IMBHs below a mass threshold might not generate a broad-line region \citep{Chakravorty2014}. For this reason, we have not considered broad-line selection in this work (but see \citealt{Cho2024}.)

Narrow emission line ratios trace the fractional contribution of AGN to star formation activity in galaxies \citep{Kewley2013}. Mapping the selection efficiencies of narrow emission line ratio diagnostic diagrams to AGN luminosity requires detailed forward modeling using photoionization codes plus an accounting for the effects of star-formation dilution, dust, and aperture effects \citep{Trump2015}. Furthermore, the ability to detect AGNs in low mass galaxies drops significantly due to the sensitivity of the line ratios to host metallicity \citep{Groves2006}. However, recent work have identified galaxies with AGN-like line ratios using metallicity-insensitive line ratios that would be classified as star-forming using the metallicity-sensitive line ratios. \citet{Polimera2022} identified these star-forming AGNs in galaxies with stellar masses as low as $M_{\star} \sim 10^{8.2} M_{\odot}$. Modeling the selection efficiency of these star-forming AGNs diagnostics to infer the occupation fraction may be worth revisiting in future work.

Mid infrared selection is very interesting due to its relative simplicity and the existence of the all sky WISE survey \citep{Stern2012,Assef2013,Hviding2022}. However, at low redshift, it suffers from substantial contamination from starburst galaxies at low luminosities \citep{Hainline2016,Satyapal2018,Asmus2020}. 

\section{Conclusions} \label{sec:conclusions}

In this paper we inferred the local nuclear black hole occupation fractions, black hole mass function, and AGN luminosity functions using a X-ray, radio, and optical variability survey data. We jointly constrained the Eddington ratio distribution functions and black hole occupation fraction, providing important observational constraints on semi-analytic model predictions (e.g., \citealt{Ricarte2018}). Our main findings are summarized below:
\begin{enumerate}
    \item We derive the complete likelihood function to infer the local black hole occupation fraction using a Bayesian generalized linear model that accounts for contamination from star-formation related emission, incompleteness, and non-detections. 
    \item We present inferred local occupation fraction limits from X-ray, radio, and optical variability. According to our analysis, the occupation fraction is at least $90$ percent at $M_{\star} = 10^8 M_{\odot}$ and at least $39$ percent at $M_{\star} = 10^{7} M_{\odot}$ (95\% credible intervals). These limits are more stringent than the occupation fraction derived from dynamical black hole mass measurements, optical line ratios in local field galaxies, NSC occupation fraction in the local volume, and previous work from the AMUSE X-ray survey. 
    
    \item We compared our results to predictions from semi-analytic model results \citep{Ricarte2018} and the analytic argument from \citet{Buchner2019}. Using the \citet{Buchner2019} argument, at least 15\% of halos with masses greater than $\sim 10^{7} M_{\odot}$ had to be seeded at high redshift. The results from the \citet{Ricarte2018} semi-analytic models favors a light seed formation scenario without excluding heavy seeding channels as well.
    
    \item Using these occupation fraction limits and assuming standard $M_{\rm{BH}}-M_{\star}$ relations, we derived a BHMF that is consistent with our observations. Our completeness and occupation corrected BHMF is higher than the BHMF derived from the $M_{\rm{BH}}-\sigma_{\ast}$ relation at the high mass end, but consistent with the BHMF derived from from tidal disruption events.
    
    \item We constrain the extremely low luminosity ends ($L_{\rm{bol}} \lesssim 10^{40}$ erg s$^{-1}$) of the X-ray, radio, and optical AGN LFs. All three of our LFs are consistent to $\sim 1\sigma$ with the $z=0$ LFs from the literature at higher AGN luminosities and indicates the AGN LFs can be extrapolated down to extremely low bolometric luminosities.
\end{enumerate}


Our work was motivated by the previous work of \citet{Miller2015}. Our work is an improvement upon the previous work in several ways. First, we assumed a smoothly broken power law ERDF that results in a much better model for the observed LFs than a log-normal ERDF. Our formalism has enabled us to self-consistently infer the ERDF and the BHOF. Finally, we extended the analysis to radio and optical datasets and used much deeper data with more detections in lower mass galaxies. This is possible because we have taken into account the parent sample completeness and upper limits for every source. As a result of our more realistic model and more detections in low-mass galaxies, our constraints on the BHOF are more than an order of magnitude more stringent than the previous results from \citet{Miller2015}.

In a follow-up paper, we will generalize the inference methodology developed here to allow the BHOF and ERDF to vary smoothly with redshift while correcting for the $L-z$ degeneracy. We will use deep field data to constrain the BHOF at higher redshifts and couple the inference to semi-analytic models of SMBH assembly. \citet{Chadayammuri2023} showed that X-ray surveys that optimize for depth and volume have greater BHOF constraining power. The ability to distinguish seeding mechanisms hinges on constraining the black hole populations in unbiased samples of nearby galaxies with $M_\star \lesssim 10^8\ M_{\odot}$. For local galaxies, the required sensitivities should be within reach of current radio facilities such as the Very Large Array, future X-ray missions such as AXIS, and the upcoming Vera C. Rubin Observatory Legacy Survey of Space and Time. Self-consistently coupling semi-analytic models to the black hole mass distributions observed with accretion probes and with LISA will be an important step toward constraining theoretical models of SMBH seeding and growth.


\section{Acknowledgments}
This research has made use of the SIMBAD database, operated at CDS, Strasbourg, France. This research has made use of data obtained from the Chandra Source Catalog, provided by the Chandra X-ray Center (CXC).

CJB is supported by an NSF Astronomy and Astrophysics Postdoctoral Fellowship under award AST-2303803. This material is based upon work supported by the National Science Foundation under Award No. 2303803. This research award is partially funded by a generous gift of Charles Simonyi to the NSF Division of Astronomical Sciences. The award is made in recognition of significant contributions to Rubin Observatory’s Legacy Survey of Space and Time. 
P.N. acknowledges support from the Gordon and Betty Moore Foundation and the John Templeton Foundation that fund the Black Hole Initiative (BHI) at Harvard University where she serves as one of the PIs.
CJB is grateful to Patrick Aleo, Alex Gagliano, Juan Guerra, Antonio Porras, Tom Maccarone, Pieter van Dokkum, and Meg Urry for useful discussion. We are grateful to the anonymous referee for a careful review that significantly improved this work.

%

\vspace{5mm}
\facilities{Chandra, XMM-Newton, eROSITA, ROSAT, VLA, PTF}
 

\software{\textsc{astropy} \citep{Astropy2022}, \textsc{emcee} \citep{Foreman-Mackey2013}, \textsc{corner} \citep{corner}, \textsc{numpy} \citep{harris2020array}, \textsc{scipy} \citep{2020SciPy-NMeth}, \textsc{matplotlib} \citep{Hunter2007}}



\appendix

\section{Derivation of Complete Log-likelihood}
 \label{sec:logL}

The observed probability distribution can be derived from a Bernoulli process with a probability $f_{\rm{d}}$, whose outcome is 1 if the black hole is detectable, and 0 if there is no detectable black hole present. The remainder of the terms simply transform the Bernoulli outcome (frequentist view) to luminosity space via the definition of the Eddington ratio. The total observed probability distribution can then be derived from a mixture model involving this ``foreground'' Bernoulli probability density and the ``background'' model from star-formation related contamination and a completeness term as:

\begin{equation}
    p(\log L | \log M_{\star}) = 
p(I | \log M_{\star})\ \bigg[ (1 - P_{\rm{SF}})\ 
{\rm{Br}}(f_{\rm d})\ {\rm{sERDF}}(\lambda | \log M_{\star}) + P_{\rm{SF}}\ \mathcal{N}(\log L - \log L_{\rm{SF}} | s^2_{\rm{SF}}) \bigg],
\end{equation}

\begin{equation}
 = 
p(I | \log M_{\star})\ \bigg[ (1 - P_{\rm{SF}})\ 
\bigg( f_{\rm{d}}\ {\rm{sERDF}}(\lambda | \log M_{\star}) +
(1-f_{\rm{d}})\ \delta(\log L + 9999) \bigg) \\ + P_{\rm{SF}}\ \mathcal{N}(\log L - \log L_{\rm{SF}} | s^2_{\rm{SF}}) \bigg],
\end{equation}

where Br($p$) is the Bernoulli probability density.

The standard likelihood including both detections, non-detections with upper limits is,
\begin{equation}
    \mathcal{L} = \prod_{i \in \mathcal{A}_{\rm det}} p(y_i | x_i)\ \times
    \prod_{j \in \mathcal{A}_{\rm cens}} \int^{y_j}_{-\infty} p(y_j | x_j)\ dy_j.
\end{equation} 
The log likelihood is,
\begin{equation}
    \log \mathcal{L} = \sum_{i \in \mathcal{A}_{\rm det}} p(y_i | x_i)\ +
    \sum_{j \in \mathcal{A}_{\rm cens}} \int^{y_j}_{-\infty} p(y_j | x_j)\ dy_j.
\end{equation} 
The first term is the product over the detected data set. The second term is the product over the censored data set containing the non-detections with upper limits $y_j$. The second term is marginalized over all possible values of the non-detections. After integrating over the delta function, the resulting log likelihood is,
\begin{multline}
\log \mathcal{L} \propto \sum_{i \in \mathcal{A}_{\rm det}} 
\log \biggl( \biggl\{ p(I_i=1 | \log M_{{\star},i})\ \biggl[ (1 - P_{\rm{SF}})\ f_{\rm{d}}\ {\rm{sERDF}}(\lambda_i | \log M_{{\star}, i}) + P_{\rm{SF}}\ \mathcal{N}(\log L_i - \log L_{{\rm{SF}}, i} | s^2_{{\rm{SF}}, i}) \biggr] \biggr) \biggr\} + \\
\sum_{j \in \mathcal{A}_{\rm cens}} \log \biggl( \biggl\{ p(I_j=1 | \log M_{{\star},j})\ \biggl[ (1 - P_{\rm{SF}})\ f_{\rm{d}}\ \int_{\lambda_{\rm{min}}}^{\lambda_j} {\rm{sERDF}}(\lambda | \log M_{{\star}, j})\ d\lambda + (1-f_{\rm{d}})\ + \\ P_{\rm{SF}}\ \Phi(\log L_j - \log L_{{\rm{SF}}, j} | s^2_{{\rm{SF}}, j}) \biggr] \biggr) \biggr\},
\end{multline} 
where $\Phi(\mu | s^2)$ is the cumulative distribution function of a normal distribution with mean $\mu$ and variance $s$. The convolutions (Equation~\ref{eq:erdf}) are calculated using \textsc{SciPy's} \texttt{fftconvolve} method. The remaining integrals are solved numerically using \textsc{SciPy's} \texttt{trapezoid} method.

\section{Updated NASA-Sloan Atlas Stellar Masses}
 \label{sec:cigale}

\begin{figure*}[ht!]
\centering
\includegraphics[width=0.5\textwidth]{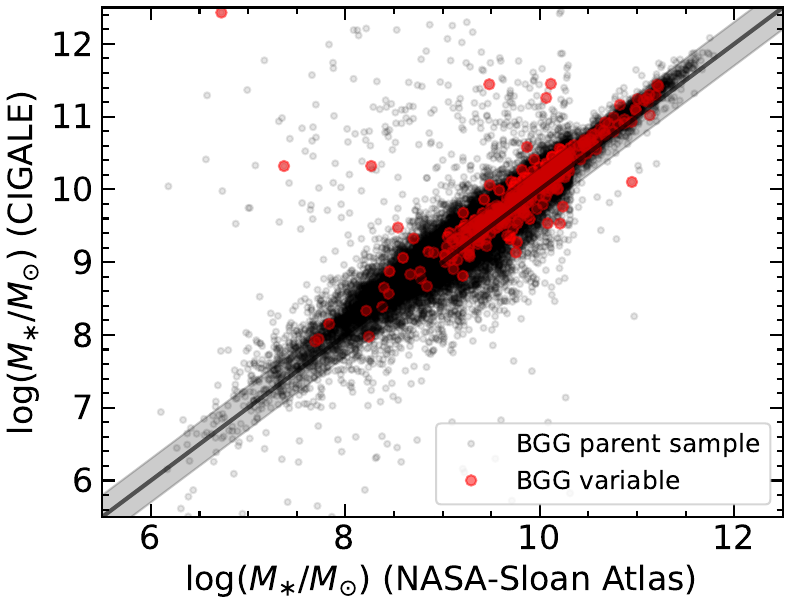}
\caption{Stellar masses with updated redshifts estimated using \textsc{cigale} versus those from the NASA-Sloan Atlas within the PTF variability parent sample (small black circle symbols) and variable sources (large red circle symbols). We use our updated \textsc{cigale} stellar masses for the optical variability sample in our analysis.
\label{fig:massnsa}}
\end{figure*}

We re-fit the NASA-Sloan Atlas stellar masses using the \textsc{cigale} SED fitting code \citep{Boquien2019,Yang2020,Yang2022} with our updated spectroscopic redshifts from SDSS and GAMA. The SED parameters includes an AGN component and are identical to that of \citet{Burke2024hsc}. A comparison of our updated stellar masses is shown in Figure~\ref{fig:massnsa}. There is a notable excess of galaxies with low stellar masses in the NASA-Sloan Atlas stellar masses due to wrong spectroscopic redshift or anomalous NASA-Sloan Atlas colors that \textsc{cigale} is better able to fit.

\section{Results from the AMUSE survey}
\label{sec:amuse}

\begin{figure*}[ht!]
\centering 
\includegraphics[width=0.98\textwidth]{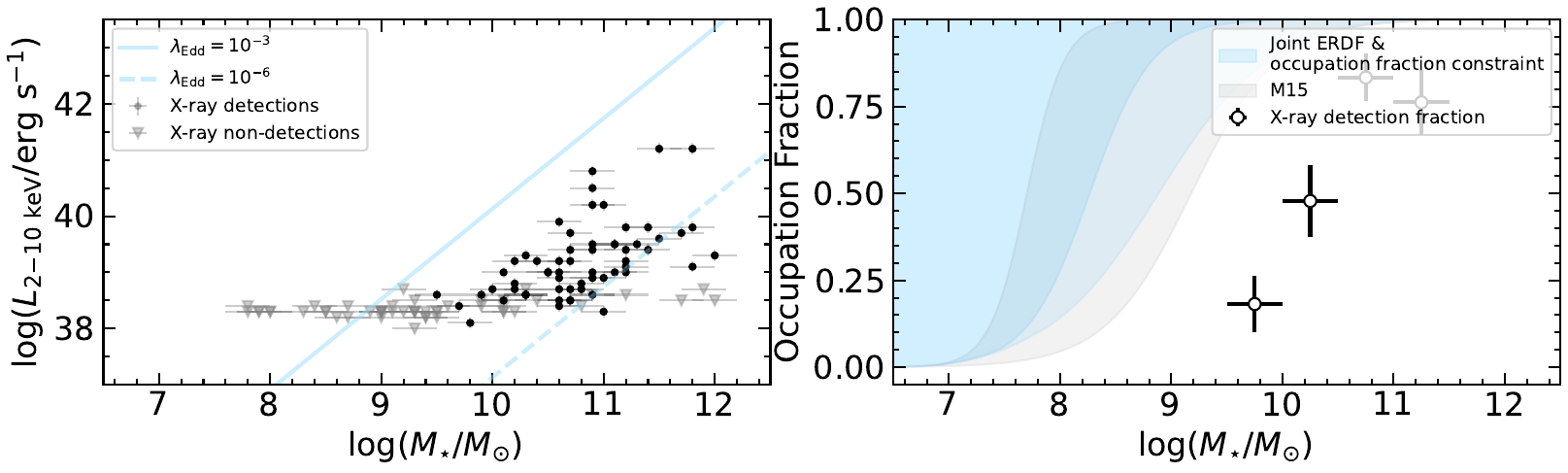}
\includegraphics[width=0.98\textwidth]{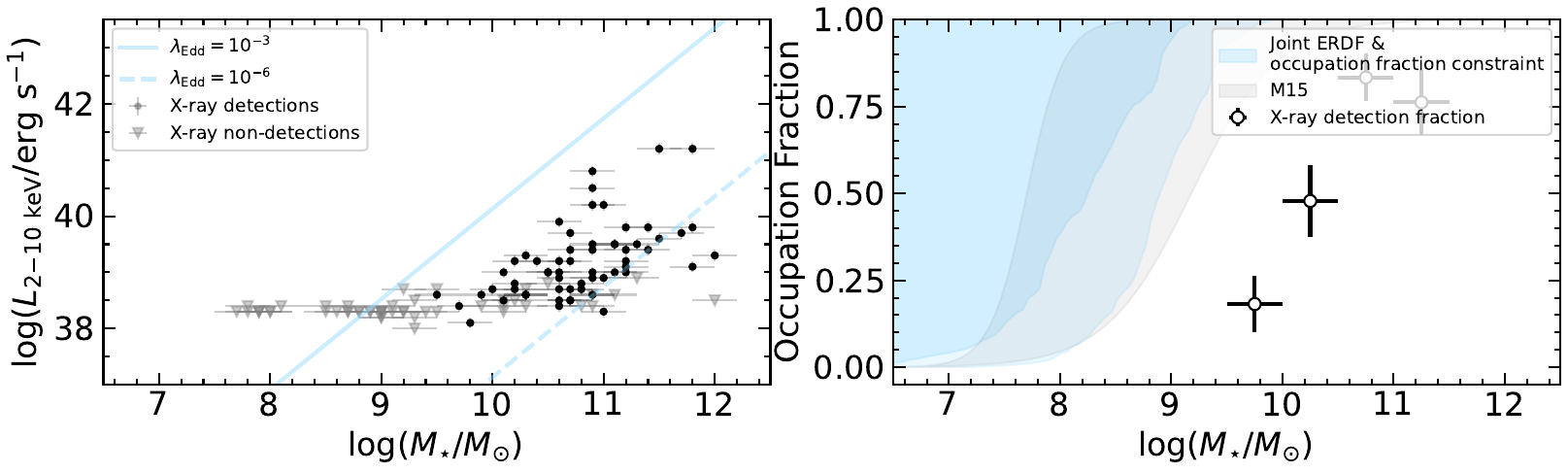}
\caption{Same as Figure~\ref{fig:inference}, but for the ``clean'' sample of the AMUSE X-ray survey of early-type galaxies from Table 1 of \citet{Miller2015}. \emph{Upper panel}: Results using the functional form of the occupation fraction as defined by \citet{Miller2015}. \emph{Lower panel}: Results using the general form of the occupation fraction defined in Equation~\ref{eq:BHOF}. The 1$\sigma$ constraints on the occupation fraction taken directly from Figure 2 of \citet{Miller2015} are shown in gray shaded region (M15). }
\label{fig:amuseinference}
\end{figure*}

We attempt to reproduce the results from \citet{Miller2015} using the AMUSE Chandra survey of early-type galaxies. We use the ``clean'' sample from their Table 1 that excludes star forming galaxies and galaxies with NSCs detected from \emph{Hubble Space Telescope} imaging. This sample of ``clean'' early-type galaxies should have minimal to no contamination from XRBs. We use the same likelihood function as before but, we adopt the functional form of the BHOF from \citet{Miller2015} of,
\begin{equation}
    f_{\rm{BHOF}}(\log M_{\star}, \log M_{50}) = 0.5 + 0.5 \tanh \bigg( 2.5^{|8.9 - \log M_{50}|} (\log M_{\star} - \log M_{50}) \bigg),
\end{equation}
where. The slope of $\sim 1.6$ ($\log L \propto \log M_{\rm{BH}} \propto 1.6 \log M_{\star}$ from the black hole mass -- stellar mass relation) is steeper than the slope of $0.79 \pm 0.12$ for the clean sample found by \citet{Miller2015}. However, \citet{Miller2015} assumed a log-normal scatter, which is inconsistent the observed ERDF. Despite these different modeling choices, the resulting constraints on the occupation fraction are very similar, with overlap between the $1 \sigma$ shaded regions. As an additional test, we repeat the analysis but using the more general functional form of the BHOF from Equation~\ref{eq:BHOF} (Richards's curve). Reassuringly, the shape of the occupation fraction constraints are quite consistent between the two functional forms of the BHOF. The results for both functional forms of the BHOF with the ``clean'' AMUSE sample are shown in Figure~\ref{fig:amuseinference}. The inferred BHOF $1\sigma$ lower limit from Figure 2 of \citet{Miller2015}, shown as the gray shaded region in Figure~\ref{fig:amuseinference}. 

Differences in the likelihood, priors, sample definition, and assumptions from XRB contributions can clearly affect the results. The \citet{Miller2015} analysis used a multiple imputation approach to model the low-mass XRB contribution. The expected XRB luminosity is estimated by scaling the XRB LF from globular clusters. Differences between the BHOF from our AMUSE sample inference and the Chandra Source Catalog sample inference can be attributed to different sample definitions. Primarily, the AMUSE sample only includes early-type galaxies. Nevertheless, the agreement demonstrated by the overlapping BHOF constraints between our results in Figure~\ref{fig:inference} and with the previous work is reassuring.

\section{Posterior Distributions from Combined Multiwavelength Analysis}
\label{sec:post}

The posterior distributions from our combined multiwavelength analysis is shown in Figure~\ref{fig:corner}. We also show partially marginalized 2D likelihood maps of $\log M_{50}$, $\delta$, $\theta$ parameters. We evaluate the likelihood on 2D grids of $\log M_{50}$, $\delta$ and  $\log M_{50}$, $\theta$. All other parameters are taken as posterior medians from the MCMC posterior (Figure~\ref{fig:corner}). The partially marginalized analysis does not properly account for covariance between parameters. The parameters $\delta$, $\theta$ are correlated through the definition of the occupation function. Unlike the MCMC sampling, this analysis is unable to constrain these shape parameters well. Hence, this figure is only taken as a simple consistency check on the inflection point in the occupation function. The analysis shows that $\log M_{50} < 8.0$ at $\sim 3\sigma$ the level, consistent with Figure~\ref{fig:occupationcombined}.

\begin{figure}[ht!]
\centering
\includegraphics[width=1\textwidth]{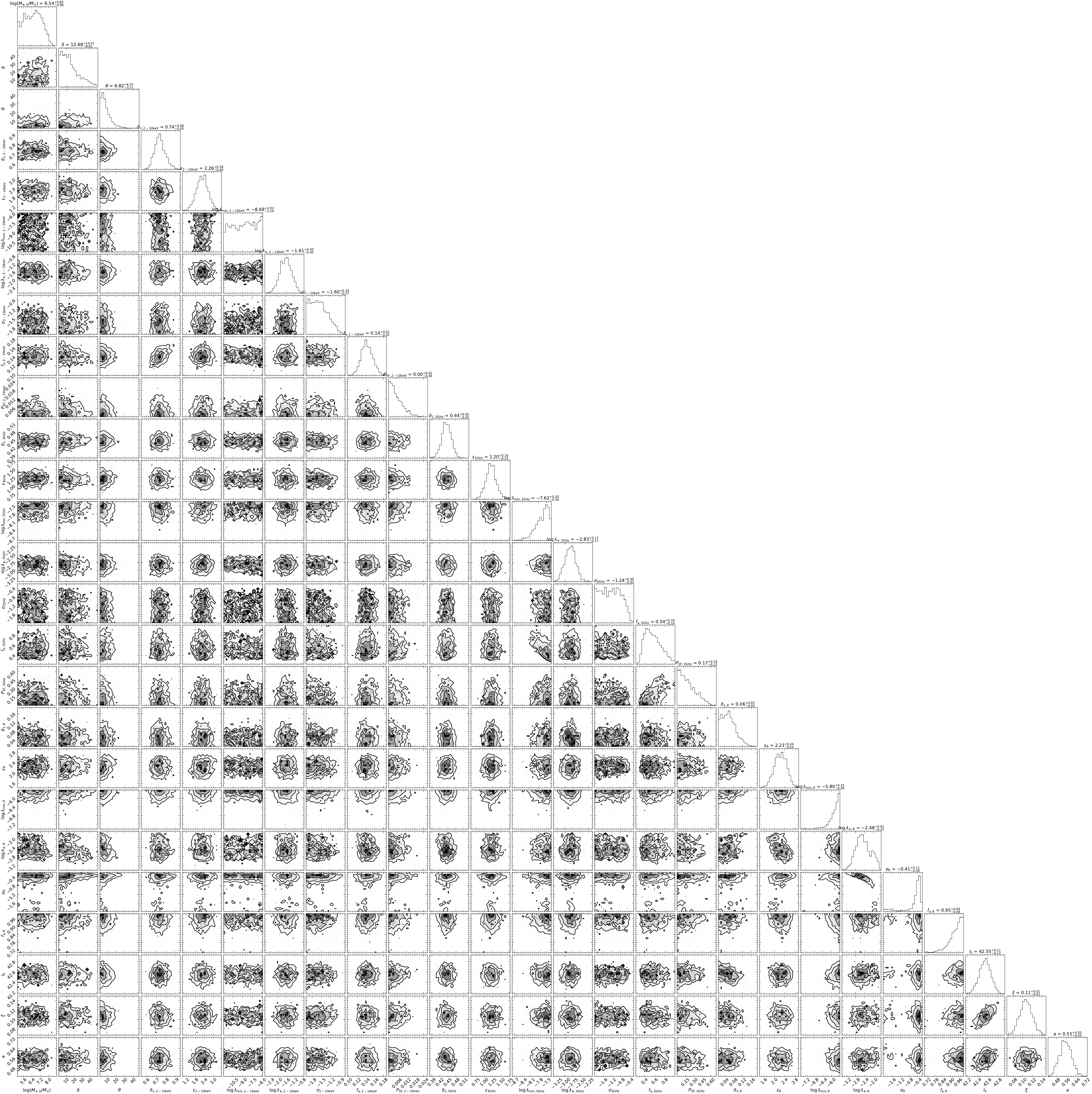}
\caption{Posterior distribution corner plot for the combined multiwavelength analysis. The posterior distributions for each parameter are shown along the diagonal. The titles above the distributions give the posterior medians and 16th and 84th percentiles. The 2D histograms show the correlations between each parameter. 
\label{fig:corner}}
\end{figure}

\begin{figure}[ht!]
\centering
\includegraphics[width=1\textwidth]{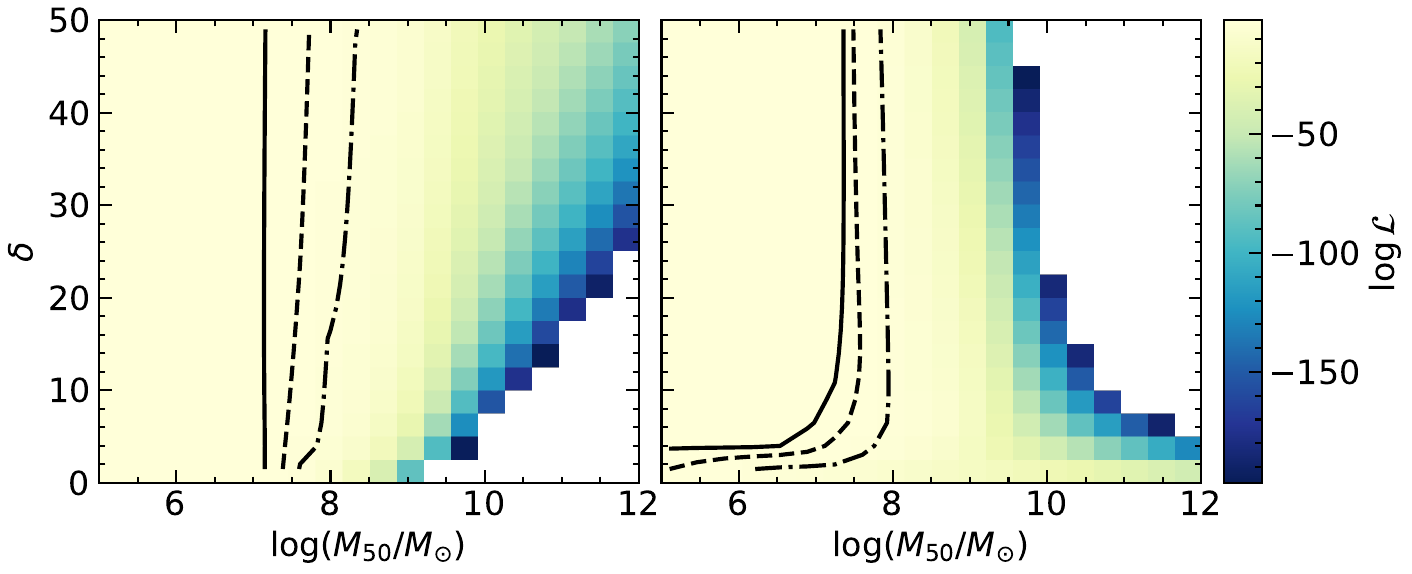}
\caption{Normalized partially marganilized likelihood surface of $\log M_{50}$ vs. $\delta$ (\emph{left}) and $\log M_{50}$ vs. $\theta$ parameters (\emph{right}). The contours are $1\sigma$ (solid), $2\sigma$ (dashed), and $3\sigma$ (dot-dashed) limits and show that $\log M_{50} < 8.0$ at $\sim 3\sigma$ the level. The white squares are points of very low likelihood. 
\label{fig:corner}}
\end{figure}


\bibliography{sample631}{}
\bibliographystyle{aasjournal}



\end{document}